\documentstyle[aps,eqsecnum,tighten,epsfig]{revtex}

\def\B{{\cal P}}
\def\A{{\cal A}}
\def\Ab{\overline{\A}}
\def\G{{\cal G}}
\def\Gb{\overline{\cal G}}

\def\C{{\cal C}}
\def\P{\tilde{\cal P}}
\def\H{{\cal H}}

\def\cyl{{\rm Cyl}}
\def\Cyl{{\rm Cyl}}
\def\av{{\rm av}}
\def\det{{\rm det}}

\def\Tr{{\rm Tr}}

\def\diag{{\rm diag}}

\def\be{\begin{equation}}
\def\ee{\end{equation}}
\def\ba{\begin{eqnarray}}
\def\ea{\end{eqnarray}}

\def\E{{\hat E}}
\def\q{{\hat q}}
\def\ab{\overline{A}}

\def\g{\gamma}
\def\k{\kappa}
\def\S{{\Sigma}}

\def\e{\epsilon}
\def\de{{\dot e}}
\def\goesto{\rightarrow}
\def\pC{\partial C}
\def\bra{\langle}
\def\ket{\rangle}

\begin{document}

\title{Quantum Theory of Geometry II: Volume operators}
\author{Abhay Ashtekar${}^{1,4}$\thanks{Electronic address: 
ashtekar@phys.psu.edu},
Jerzy Lewandowski${}^{2,3,4}$\thanks{Electronic address:
jerzy.lewandowski@fuw.edu.pl}}  
\address{$1$ Center for Gravitational Physics and Geometry\\
Department of Physics, Penn State,\\ 
University Park, PA 16802-6300, USA
} 
\address{$2$ Instytut Fizyki Teoretycznej, Uniwersytet Warszawski\\ 
ul. Hoza 69, 00-681 Warszawa, Poland
}
\address{$3$ Max-Planck-Institut f\"ur Gravitationsphysik, 
Albert Einstein Institut,\\ Schlaatzweg 1, D-14473 Potsdam, Germany}
\address{$4$ Erwin Schr\"odinger International Institute 
for Mathematical Sciences\\
Boltzmanngasse 9, A-1090 Vienna, Austria
}
\maketitle

\begin{abstract}
A functional calculus on the space of (generalized) connections was
recently introduced without any reference to a background metric.  It
is used to continue the exploration of the quantum Riemannian
geometry.  Operators corresponding to volume of three-dimensional
regions are introduced rigorously. It is shown that there are two
natural regularization schemes, each of which leads to a well-defined
operator. Both operators can be completely specified by giving their
action on states labelled by graphs.  The two final results are
closely related but differ from one another in that one of the
operators is sensitive to the differential structure of graphs at
their vertices while the second is sensitive only to the topological
characteristics.  (The second operator was first introduced by Rovelli
and Smolin and De Pietri and Rovelli using a somewhat different
framework.) The difference between the two operators can be attributed
directly to the standard quantization ambiguity. Underlying
assumptions and subtleties of regularization procedures are discussed
in detail in both cases because volume operators play an important
role in the current discussions of quantum dynamics.

\end{abstract} 

\section{Introduction}
\label{sec 1}

Riemannian geometry provides the mathematical framework for general
relativity and other modern theories of gravity. One therefore expects
that a non-perturbative formulation of quantum gravity would require a
corresponding quantum theory of geometry, and, at the same time,
provide pointers for constructing this theory. Familiar Riemannian
geometry would then emerge only as an approximation on coarse graining
of the semi-classical states.  In a specific non-perturbative approach
based on canonical quantization, these expectations are being borne
out in detail.  The goal of this series of papers is to present the
resulting quantum theory of geometry.  Basic techniques were developed
in \cite{1} and applied to the problem of constructing area operators.
The purpose of this paper is to carry out a similar construction of
volume operators. (The volume operator we derive in the main text was
introduced in \cite{7}.)

Let us begin with a brief summary. In the canonical quantization
approach used here, the configuration variable is an $SU(2)$
connection $A_a^i(x)$ on a three-manifold $\S$. (Indices $a,b,c, ...$
refer to the tangent space of $\S$ and indices $i,j,k, ....$ are the
$su(2)$ Lie-algebra indices.)  The momentum variable is a vector
density $E^a_i(x)$ with values in the $su(2)$ Lie algebra (or,
equivalently, a (pseudo) two-form $e_{abi} := \eta_{abc}
E^c_i$, where $\eta_{abc}$ is the Levi-Civita pseudo density). In the
quantum theory, then, one is naturally led to consider the space $\Ab$
of (suitably generalized) connections on $\S$ as the (quantum)
configuration space. To obtain the Hilbert space $\H$ of quantum
states and geometric operators thereon, one needs a functional
calculus on $\Ab$ which also does not refer to a fiducial metric (or
any other background field).

The necessary tools were developed in a series of papers by a number
of authors \cite{2,3,4,5,6,7,8,9,10,11}. (Much of the motivation
for this work came from the `loop representation' introduced earlier
by Rovelli and Smolin \cite{13}.)  It turns out that $\Ab$ admits a
natural diffeomorphism invariant measure $\mu^o$ and the Hilbert space
$\H$ can be taken to be the space $L^2(\Ab, d\mu^o)$ of
square-integrable functions on $\Ab$ \cite{3,4,5,6,7,8}. Physically,
$\H$ represents the space of kinematic quantum states, i.e., the
quantum analog of the full phase space.  Using the well-developed
differential geometry on $\Ab$ \cite{7}, one can then define
physically interesting operators on $\H$. In particular, one can
introduce, in a systematic manner, operator-valued distributions $\E$
corresponding to the triads \cite{1}. As in classical Riemannian
geometry, these are the basic objects in the quantum
case. Specifically, the idea is to construct geometric operators
--e.g., those corresponding to area, volume and length-- by
regularizing the appropriate products of these triad operators.  

As remarked above, triads --being density weighted-- can be naturally
thought of as pseudo two-forms $e_{abi}$. To obtain phase space
functions, it is natural to smear them against Lie-algebra-valued test
fields $f_i$ with support on {\it two-dimensional surfaces}. Can the
corresponding quantum operators be well-behaved? The answer is not
apriori obvious: from Minkowskian quantum field theory, one would
expect that well-defined operators will not result unless they are
smeared in (at least) three dimensions. Somewhat surprisingly,
however, for triads the answer is in the affirmative. More
generally, in this approach to quantum geometry, there is a remarkable
synergy between geometry and analysis: in the regularization
procedure, well-defined operators result when n-forms are integrated
on n-manifolds. Thus, the operators that code information in
connections are holonomies $\hat{h}[\alpha]$, obtained by integrating
the connection one-forms along one-dimensional curves. The two-forms triad
operators are naturally regulated through a two-dimensional
smearing. This feature is deeply intertwined with the underlying
diffeomorphism invariance of the theory. By contrast, in the quantum
theory of Maxwell fields in Minkowski space-time, for example, one
smears both connection one-forms and electric field two-forms in
three dimensions, using the geometrical structures made available by
the background metric.

In \cite{1}, square-roots of appropriate products of the triad
operators were regularized to obtain area operators $\hat{A}_S$
associated with two-dimensional surfaces $S$ without boundary. These
are the quantum analogs of the area functions $A_S$ defined on the
classical phase space. We now wish to discuss the volume operator
$\hat{V}_R$ associated with a three-dimensional region $R$ ---the
quantum analog of the function $V_R:= \int_R d^3x \, \sqrt{|{\rm
det}E|}$ on the classical phase space. Since $V_R$ is a rather
complicated, non-polynomial function of the triads, as one might
imagine, the issue of regularization is quite subtle. Indeed, the
problem turns out to be considerably more complicated than that for
area operators. In particular (for the continuum theory) it appears
that, so far, no regularization scheme has appeared in the literature
which has the following rather basic property: If we denote by
$\hat{V}_R^\epsilon$ the regulated version of the volume operator, then (in
the Hilbert space topology) the limit $\lim_{\e\rightarrow 0}
(\hat{V}_R^\e \cdot \Psi)$ should exist for a dense subset of states
$\Psi\in\H$. We will rectify this situation.  In the process, we will
also spell out the underlying assumptions and point out some other
subtleties that are often overlooked. Finally, in our treatment, the
limit is achieved at a finite stage, i.e., for all
$\epsilon\le\epsilon_\Psi$. This property makes our regularization of
volume operators consistent with that of the Hamiltonian constraint;
there is thus a uniform scheme that is applicable for all operators of
physical interest.%
\footnote{For a further discussion of this point, see the beginning of
the Appendix.}

In the main body of the paper, we will discuss a regularization along
the lines that led us to area operators in \cite{1}. (The final form
of this volume operator was reported in \cite{7,14,15} and its
restriction to a lattice theory was discussed by Loll \cite{16}.  The
operator was also studied by Thiemann in \cite{Thomas2}.) We will see
that this operator is sensitive to the differential structure on the
three-manifold $\S$.  In the appendix, we will consider a different
regularization scheme which is based on a construction given by
Rovelli and Smolin \cite{17} in the loop representation. (The final
closed-form expression of that operator was introduced by De Pietri
and Rovelli in \cite{18}, and, in the framework discussed here, in
\cite{15}.) In a certain sense, this operator is not sensitive to the
differential structure of $\S$.  We will see that both operators can
be constructed through systematic regularizations.  They are
well-defined, self-adjoint operators on $\H$ with purely discrete
spectrum. The actual expressions of the two operators are rather
similar and the difference between them can be interpreted simply as a
`quantization ambiguity'.  Nonetheless, in various applications, e.g.,
to quantum dynamics \cite{19}, they can lead to important
differences. To get a deeper understanding of the relation between
them, one needs to further analyze their properties --e.g., their
relation to the area and length \cite{length} operators. Such a
systematic analysis has begun only recently.

The paper is organized as follows. Section \ref{sec2} is devoted to
preliminaries.  Regularization leading to the first operator is
discussed in detail in Sections \ref{sec3} and \ref{sec4} and, using
techniques introduced by De Pietri, the regularization leading to the
Rovelli-Smolin operator is discussed in the Appendix.  Some
properties of the volume operator are discussed in Section
\ref{sec5}. Section \ref{sec6} summarizes the main results and
compares the regularization procedures.

For simplicity, in the main discussion, we have set $c=1$, $8\pi G =1$
and $\hbar =1$ and chosen the real connection $A_a^i$ to be $A_a^i =
\Gamma_a^i - K_a^i$ where $\Gamma_a^i$ is the spin connection
compatible with the triad $E^a_i$ and $K_a^i$ is the extrinsic
curvature. As pointed out by Immirzi \cite{immirzi} using earlier work
of Barbero \cite{barbero}, unitarily inequivalent quantum theories
result if one begins with the canonical pair ${}^\gamma\!A_a^i =
\Gamma_a^i - \gamma K_a^i$, ${}^\gamma\!E^a_i = (1/\gamma)\, E^a_i$,
where $\gamma$ is a non-zero real parameter. Thus, in the main
discussion, we will work with the $\gamma =1$ sector. In Section
\ref{sec4.4}, we will restore $c$, $\hbar$ and $G$ and also state the
main result in any Immirzi sector.

\section{Preliminaries}
\label{sec2}

In this section, we briefly recall the mathematical ideas that
underlie the present approach to quantum Riemannian geometry. This
discussion will also serve to fix notation. It turns out that some
diversity has arisen in viewpoints and conventions in the recent
literature on non-perturbative quantum gravity.  To remove potential
confusion, therefore, the corresponding issues will be discussed in
detail.

Fix an orientable, analytic%
\footnote{In this work the assumption of analyticity is not essential;
we make it for simplicity since it allows us to use previous results
\cite{1,2,3,4,6,7,8,9} directly.  Our constructions can be made to go
through for smooth manifolds and graphs. This point is discussed at
the end of section \ref{sec4}.}
three-manifold $\S$ and a principal $SU(2)$ bundle $B$ over $\S$. Our
configuration space $\C$ will consist of smooth connections on
$B$. Since all $SU(2)$ bundles over three-manifolds are trivial, we
can fix a trivialization and regard each connection $A$ on $B$ as an
$su(2)$-valued one-form $A_a^i$ on $\S$, where $a$ is the form index,
and $i$, the Lie-algebra index. (We will not specify boundary
conditions on fields because they are irrelevant for the issues we
wish to discuss here.) The `conjugate momenta' are {\it
non-degenerate} vector densities $E^a_i$ of weight one --- or,
equivalently, pseudo two-forms $e_{abi}:= \textstyle \eta_{abc}E^c_i$
--- with values in $su(2)$, where $\eta^{abc}$ is the Levi-Civita
pseudo density.  Thus, the action
$$\int_\S d^3x\, E^a_i \delta{A}_a^i \,\, \equiv \,\,
\int_\S \delta A^i \wedge e_{i}$$
of the cotangent vector $E^a_i$ on a tangent vector $\delta{A}_a^i$ is
invariant under the change of orientation of $\S$. Although they are
density weighted, for brevity, the $E^a_i$ will be referred to simply 
as {\it triads}.
  
Riemannian geometry of the three-manifold $\S$ is coded in the momenta
$E^a_i$. To see this, note first that given vector densities $E^a_i$,
we can define a triplet of vector fields $e^a_i$ via:
\be\label{E} 
e_i^a\ =\ {E^{a}_i\over \sqrt{|\det\, E|}} , \ee 
where $\det \, E$ stands for the determinant of the matrix
$(E^{ia})$ with $i,a=1,2,3$. Note that the phase space contains frame
fields $e^a_i$ with {\it both} orientations. Given these fields, we
can just define a contravariant, positive definite metric $q^{ab} :=
e^a_i e^b_j k^{ij}$ where $k=-2\Tr$ is the Cartan-Killing metric on
$su(2)$. If we denote by $q$ the determinant of $q_{ab}$ (the inverse
of $q^{ab}$), we also have: $E^a_i = \sqrt{q} e^a_i \equiv |{\rm det} e|
e^a_i$. In terms of these Riemannian structures, the volume of a
region $R$ (covered, for simplicity, by a single chart) is given by:
\be
V_R:= \int_R d^3x \sqrt{q} \equiv \int_R d^3x \sqrt{|\det E|}\, .
\ee

As noted in the Introduction, in quantum theory, one is naturally led
\cite{2,3} to consider the space $\Ab$ of (suitably generalized)
connections as the configuration space. Thus, the Hilbert space $\H$
of (kinematic) quantum states is given by $\H = L^2(\Ab, d\mu^o)$
where $\mu^o$ is a natural diffeomorphism invariant measure on $\Ab$
\cite{3,4,5,6,7,8}. $\H$ contains a dense subspace $\cyl$ of
`cylindrical functions' which turns out to be especially useful. These
are constructed as follows. Each element $\ab$ of $\Ab$ assigns to any
analytic path $p$ in $\S$, an element $\overline{A}(p)$ of $SU(2)$,
(which can be regarded as the `holonomy' of the generalized connection
$\ab$) \cite{7}. Fix a graph $\g$ with a finite number (say $N$) of
edges%
\footnote{By a graph $\g$ we mean a finite set $\{e_1,...,e_N\}$ of
oriented one-dimensional, analytic sub-manifolds of $\Sigma$ such that
each $e_I$ has a (two point) boundary and for $I_1\not=I_2$ the
intersection $e_{I_1}\cap_{I_2}$ is contained in the boundaries. The
elements $e_I$ of a graph are called edges, the points in their
boundaries the vertices.}
$e_I$, $I=1,...,N$, and a complex-valued function 
$\psi: [SU(2)]^N \rightarrow {\bf C}$ on
$[SU(2)]^N$. Then, we can define a function on $\Ab$ as follows:
\be \label{cylin}
\Psi_{\g} (\ab)\, = \, \psi(\ab(e_1), ... ,\ab(e_N)) \ee
(Strictly, the function on the left side should be written as
$\Psi_{\g,\psi}$.  However, for notational simplicity, we will omit
the subscript $\psi$.)  Note that $\Psi_{\g}$ `knows' only about what
the connection $\ab$ does on the $N$ edges of $\g$; it depends only of
a finite number of `coordinates' on $\Ab$. Therefore, following
standard terminology, $\Psi_{\g}$ are called cylindrical
functions. The space of cylindrical functions defined by a fixed graph
$\g$ is infinite dimensional but `rather small' in the sense that it
can be thought of as the space of quantum states of a system with only
a finite number of degrees of freedom. However, as we vary $\g$
through all finite graphs on $\S$, we obtain a space of functions on
$\Ab$ which is very large. This is the space $\cyl$. Since it is dense
in $\H$, we can first define physically interesting operators on
$\cyl$ and then consider their self-adjoint extensions. This strategy
turns out to be especially convenient in practice.

Of special interest to us are `angular momentum like' operators
$J^i_{x,e}$, associated with a point $x$ in $\S$ and an edge $e$ which
begins at $x$, where, as before, $i$ is an $su(2)$ index.  Given
$\Psi_\g$ which can be regarded as cylindrical ---i.e., represented as
in (\ref{cylin})--- with respect to a graph which does not contain a
segment of $e$ with $x$ as one of its end points,
\be (J^i_{x,e}\, \Psi_{\g})(\ab) \, = \, 0.\ee
If on the other hand $\g$ has an edge $e_J$ which shares a finite
segment with $e$ originating at $x$, without loss of generality, we
can assume that $x$ is a vertex of $\g$. Then,
\begin{figure}
\centerline{\epsfig{figure=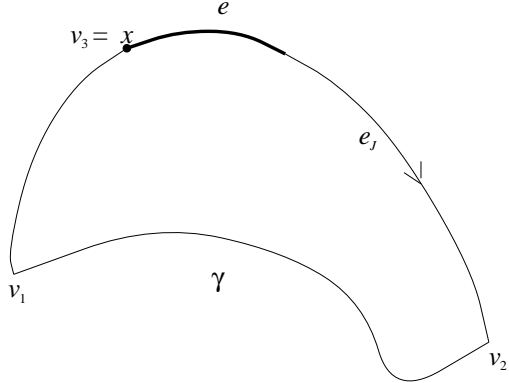,width=2in,angle=270}}
\caption{Illustration of the action of $J^i_{xe}$ operator on
$\Psi_\g$. }
\end{figure}
%
%
\ba \label{Jop}
(J^i_{x,e}\, \Psi_{\g})(\ab) \, &=& \, i(\ab(e_J)\tau^i)^A_B \,
\frac{\partial \psi} {\partial (\ab (e_J))^A_B} \nonumber\\
&=& \, i\frac{d}{dt}[\psi(\ab(e_1), ..., \ab(e_J){\rm exp}(t\tau^i),
...\ab(e_N)]\ea
if $e_J$ is outgoing at $x$.%
\footnote{The operator on the right side of (\ref{Jop}) is just the
(Lie derivative by the) left-invariant vector field $L^i$ of the copy
of $SU(2)$ associated with the $J$-th edge. It is in this sense that
$J^i_{x,e}$ are angular momentum-like operators. If $x$ is the
end-point of the edge $e_J$ rather than the beginning-point, $L_i$ 
should be replaced by the right invariant vector field $-R^i$. For 
details, see Eq. (3.9) - (3.11) in Ref. \cite{1}.}
Here, $\tau^i$ are $su(2)$ matrices, satisfying 
$$ 
-2 {\rm Tr}\, \tau^i\tau^j = k^{ij},
\quad {\rm and} \quad [\tau^i,\, \tau^j] = \epsilon^{ijk}\tau^k$$
where $k_{ij}$ is the Killing form on $SU(2)$.  

These operators are well-defined: If an element of $\cyl$ is
cylindrical with respect to two distinct graphs $\g$ and $\g'$,
$\Psi_{\g'} = \Psi_{\g}$, then $ J^i_{x,e} \Psi_{\g'} = J^i_{x,e}
\Psi_{\g}$. Denote by $\cyl^{(1)}$ the space of all cylindrical
functions on $\Ab$ for which $\psi$ in (\ref{cylin}) is $C^1$. One can
show that $J^i_{x,e}$ is essentially self-adjoint on the domain
$\cyl^{(1)}$. (For general results on essential self-adjointness of
such operators, See \cite{7}.) Next, given a point $x\in\S$ let us
introduce an equivalence relation for the edges sharing $x$ as one of
the ends: $e\sim e'$ if and only if there is a neighbourhood of $x$ on
which $e$ and $e'$ overlap modulo orientation.  Then, $J^i_{x,e} =
J^{i'}_{x',e'}$ if and only if $x=x',\, i=i'$ and $e\sim e'$.
Finally, the commutation relations between these operators are given
by:
\be\label{Jcom} [J^i_{x,e},\, J^j_{x',e'}] = i \delta_{x,x'}\, 
\delta_{[e],[e']}\, \epsilon^{ij}{}_{k} J^k_{x,e}\, , \ee
where $[e]$ denotes the equivalence class of edges defined by the
above equivalence relation and $\epsilon^{ij}{}_{k}$ are the structure
constants of $su(2)$. In what follows, $[e]$ will be referred to as
the {\it germ} of an edge $e$ starting at $x$.

Finally, let us recall from \cite{1} the definition of the smeared
triad operators, the quantum analogs of the classical expressions
$E^i_S:= \textstyle{1\over 2} \int_S\eta_{abc}E^{ai} dx^a\wedge dx^b$,
where $S$ is an analytic two-surface without boundary.  (Note that the
action of $E^i_S$ depends on the relative orientation of $S$ and
$\Sigma$ which naturally defines the notion of `up' and `down' with
respect to $S$). A careful regularization leads to the following
result:
\be\label{hE2} 
\E^i_{S}\ =\ {1\over 2}\sum_{x\in S}\,\, \sum_{[e]} \k_S([e])
J^i_{x,e}, \ee 
where the first sum ranges over all the points $x$ of $S$ and, for
every $x$, $[e]$ runs through the set of germs starting at $x$,
and where 
\be \label{kappa1}
\k_S([e]) \ =\ \cases{1, &if $e$ lies above $S$\cr
              -1,&if $e$ lies below $S$,\cr
               0,& otherwise.\cr}
\ee
The sums in (\ref{hE2}) are infinite. However, when we act the right
side on any cylindrical function, the result is well-defined because
the sums have only a finite number of non-vanishing terms.  Given a 
function $\Psi\in \cyl$ it is convinient to represent it by (\ref{cylin}) 
using a graph $\g$ such that  every isolated intersection point 
between the range of $\g$ and $S$ is a vertex of $\g$. 
Then, the action of $\E^i_S$ on $\Psi = \Psi_\g$ reads
\be \label{sto}
\E^i_{S}\,\Psi_{\g}\ 
=\ {1\over2}\sum_v\sum_{I}\k_S([e_{I}])\,  J^i_{v,e_{I}}\,\Psi_\g \ee 
where $v$ runs through the vertices of $\g$ contained in $S$, and $I$  
through the (labels of) the edges of $\g$ intersecting $v$. The operator 
$$ \E^i_{S}\ :\, \cyl^{(1)}(\Ab)\ \rightarrow\ \cyl^{(0)}(\Ab) $$
turns out to be essentially self-adjoint \cite{1}. 

This concludes our discussion of mathematical preliminaries. In the
remainder of the paper, the operators $J^i_{x,e}$ and $\E^i_S$ on
$\cyl^{(1)}$ will be used repeatedly.

\section{`Internal' or `intrinsic' regularization} 
\label{sec3}

We can now introduce the first regularization of the volume operators.
This discussion will be divided in to 3 parts. In the first, we
consider the classical theory and introduce an approximate expression
of the volume functional by dividing the region under consideration
into small cells.  In the second, we take over this `regulated'
expression to the quantum theory. In the last part, we let the cells
shrink and show that the limit yields a well-defined operator.
However, this operator carries a memory of the background structure
(namely, coordinates) introduced in the intermediate steps. To obtain
a covariant operator, we need to `average' over the relevant
background structures. The averaging procedure is carried out in
Section \ref{sec4}.

Fix an open region $R$ in $\S$. We wish to construct the operator
$\hat{V}_R$ corresponding to the function 
\be V_R (E):= \int_R d^3x\, |{\rm det}\, E|^{1\over 2} \ee
on the classical phase space. For the regularization procedure, it
will be necessary to assume that $R$ can be covered by a single
coordinate system ($x^a$). However, it turns out that this assumption
is not overly restrictive.  To see this, note first that given any
region $R$, we can cover it by a family ${\cal U}$ of neighborhoods
such that each $U\in {\cal U}$ is covered by a single coordinate
system. Let $(\phi_U)_{U\in {\cal U}}$ be a partition of unity
associated to ${\cal U}$. Then, if we set
$$ V_{R,{\phi_U}} = \int_R  d^3x\, 
\phi_U |{\rm det}\, E|^{1\over 2}$$
we have
\be\label{unity}
 V_R = \sum_{U\in{\cal U}} V_{R, \phi_U}\, ,
\ee
where the final answer is independent of the specific choice of the
partition of unity. It turns out that the same reasoning holds for the
quantum operator $\hat{V}_R$. Hence, it will suffice to define the
volume operators for regions $R$ which {\it can} be covered by a single
coordinate system. 

\subsection{Classical basis for the regularization procedure}
\label{sec3.1}

Fix global coordinates $x^a$ in a neighborhood in $\S$ containing $R$
and cover it with a family $\C$ of closed cubes whose sides are
parallel to the coordinate planes. (If a cube $C$ is not
contained in $R$, consider only its intersection with $R$.)  Two
different cells can share only points in their boundaries.  {\it
Within} each cell $C\in \C$ consider an ordered triple
$s=(S_1,S_2,S_3)$ of oriented two-surfaces (without boundary) defined
by
$$  x^a= {\rm const}^a\ \ \ a=1,2,3, $$
which intersect in the {\it interior} of $C$ (and whose orientation is
induced by that of the coordinate axes).  Such a family of pairs
$(C,s)$, of cells and triples of surfaces $s$ (`dual' to $C$), will be
called {\it a partition of} $R$ and denoted by $\B$.  Given $\B$, for
every $C\in \C$ and $s=(S_1,S_2,S_3)$ we define a functional on the
classical phase space
\be\label{q} 
q_{C}[E] \ :=\ {1\over 3!} \e^{ijk} \,\eta_{abc}\,\,
E^i_{S_{a}} E^j_{S_{b}}E^k_{S_{c}} \ee 
where, as before, $E^i_{S}= {1\over 2}\int_S e_{ab}^i dx^a\wedge dx^b$ 
is the `$S$-smeared triad'. Clearly, $|q_C(E)|/L_C^6$,
where $L_C$ is the (coordinate) size of $C$, approximates the  determinant 
$q$ of the metric $q_{ab}$ 
(defined by the triad $E^a_i$) at any internal point of the cell-cube $C$.
This approximate expression of $q$ naturally provides an approximate
expression $V_R^{\B}[E]$ of the volume of the region $R$, associated
with the partition $\B$ 
\be\label{apprvol} 
V_R^{\B}[E]\ :=\ \sum_{C\in \C}\sqrt{|q_{C}[E]|}.  \ee 
Indeed, if we assume that for some $\e>0$  $L_C$
is bounded from above by $\e$ (i.e.,  $L_C\ <\ \e$ for every cell) and
for each $\e$ we fix a partition $\B_\e$ as above, then for every
triad $E$ we have:
$$
V_R^{\B_\e}[E]\, \rightarrow \, V_R(E) \ \ \ {\rm as}\ \ \ \e
\rightarrow\ 0.  $$

Thus, in the classical theory, the phase space function $V_R(E)$ can
be expressed in terms of the two-dimensionally smeared triads.  Since
we already have the quantum operators corresponding to smeared triads,
in the next sub-section we will be able to construct regulated quantum
operators $\q_C$. The removal of regulators will however be much more
subtle. In particular, we will have to introduce a certain `averaging'
procedure. We will conclude this sub-section by justifying this
procedure from a classical perspective.

Let ${\cal S}$ denote an $n$-parameter family of coordinate systems,
containing and smoothly related to $x^a$. (As we will see in Section
\ref{sec4.1}, in quantum theory, one is led to a specific ${\cal S}$.
In the classical theory, however,we can keep ${\cal S}$ general.)  Let
us label the points of ${\cal S}$ by $n$ parameters, say
$\theta^A$. Then, repeating steps given above, we obtain $n$ parameter
families of regulated functionals of triads, $q_C^{\theta}[E]$ and
$V_R^{\B_\e(\theta)}[E]$. For each $\theta$, we obtain
$V_R^{\B_{\e}(\theta)}[E]$ such that $V_R^{\B_{\e}(\theta)}[E]\,
\rightarrow \, V_R(E)$ as $\e \rightarrow\ 0$. Let us assume that the
family of coordinate systems is such that the convergence is uniform
in $\theta$. Then, we can introduce a (rather trivial) averaging
procedure as follows. Given a normalized function $\mu(\theta)$ on
${\cal S}$, (i.e., $\int_{\cal S}d^n \theta \mu(\theta) =1$), set
\ba\label{qav}
q_{C}^{\av} [E] &:=& \int_{\cal S} d^n(\theta) \mu(\theta) q_C^{\theta}
\nonumber\\
V_{R}^{\av}[E]  &:=&  \sum_{C\in \C}\sqrt{|q_{C}^{\av}[E]|}.  
\ea
Then, $V_R^{\av} [E]\, \rightarrow \, V_R(E)$ as $\e\rightarrow\ 0$.
Like all other steps in the regularization procedure, averaging is of
course unnecessary in the classical theory. However, we will see in
Section \ref{sec4} that it plays an important role in the quantum
theory.

This regularization is called `internal' because the regulated volume
functional is expressed in terms of triads which are smeared over two
surfaces passing through the {\it interior} of cells
(see the condition $i)$ in Section \ref{sec3.3}). As a result, the
final operator will turn out to be sensitive to the relation between
tangent vectors to the edges at vertices of graphs, i.e. to the {\it
intrinsic} structure of the graph at vertices.  In the Appendix, we
discuss an `external' regularization where the regulated volume
functional is expressed in terms of triads smeared on the {\it
boundary} of cells. This expression does not depend on the details of
what happens {\it inside} any cell whence the resulting quantum
operator is sensitive only to the {\it extrinsic} structure which can
be registered on the boundaries of cells surrounding vertices.

\subsection{Regularized quantum operators}
\label{sec3.2}

The regulated volume $V_R^{\B}$ of (\ref{apprvol}) depends on the
classical phase space variables only through (two-dimensionally) smeared
triads.  Since we already have the quantum analogs $\E^i_{S}$ of these
(see (\ref{sto})), it is straightforward to define the regulated
volume operator. The operator $\q_C$ corresponding to (\ref{q}) is
given simply by:
\ba\label{hq1}
\q_{C}\, &=&\, {1\over 3!} \e_{ijk} \eta_{abc} 
\E^i_{S_a}\E^j_{S_b}\E^k_{S_c} \nonumber\\
&=& \, {1\over 48}\e_{ijk}\eta_{abc}\sum_{x_1\in S_a}\sum_{x_2\in S_b}
\sum_{x_3\in S_c}\sum_{[e_1],[e_2],[e_3]}
\k^a([e_1])\k^b([e_2])\k^c([e_3])\, J^i_{x_1,e_1}J^j_{x_2,e_2}
J^k_{x_3,e_3},
\ea
where we denoted 
\be\label{det1}
\k^d([e]):=\k_{S_d}([e]),
\ee
and where $[e_r]$ runs
through the set od germs starting at $x_r$, $r=1,2,3$.  
As in Section \ref{sec2}, although infinite sums are involved, the
action of this operator on cylindrical functions is well-defined
because the result has only a finite number of non-zero terms.  To
define the regulated volume operator $\hat{V}_R^{(\B)}$, we need to
take the absolute value and  square-root of $\q_{C}$. For this, it is 
necessary to show that $\q_{C}$ is a  self-adjoint operator. Now, we know that
each $J^i_{x,e}$ is an essentially self-adjoint operator. Furthermore,
whenever $[e_1]=[e_2]$, we have $\eta_{abc}\k^a([e_1])\k^b([e_2])
\k^c([e_3])=0$.  Therefore, the products on the right side of
(\ref{hq1}) contain operators associated with {\it distinct}
edges. These operators commute. Hence, the sum contains only products
of commuting essentially self-adjoint operators. It is easy to verify
that the right side of (\ref{hq1}) is therefore an essentially
self-adjoint operator on the domain $\cyl^{(3)}$ of $C^3$ cylindrical
functions. Hence, we can take its self-adjoint extension and a
well-defined regulated volume operator via:
\be \label{hv1}
\hat{V}_R^{\B}\:=\ \sum_C |\q_C|^{1\over 2}.  \ee
By construction, $\hat{V}_R^{\B}$ is a non-negative self-adjoint
operator. This is the quantum analog of the approximate volume
functional $V_R^{\B}$. It depends on our choice of partition $\B$ of
the region $R$.

\subsection{Removing the regulator}
\label{sec3.3}

In the classical theory, it is straightforward to remove the
regulator. We can begin with any partition $\B$ and let the cells $C$
shrink in any smooth fashion; in the limit $\e \rightarrow 0$, we have
$V_R^{\B_\e}(E)\, \rightarrow \, V_R(E)$. In the quantum theory, on
the other hand, the limiting procedure involves certain subtleties.
More precisely, now one has to `stream-line' the limiting procedure by
specifying appropriate restrictions on how the partition $\B$ is to be
refined as $\e$ tends to zero. Note however that such subtleties are a
common-place in quantum field theory. For example, in interacting
scalar field theories in low dimensions one generally has to remove
the regulators in a specific order and/or take limits keeping certain
ratios of cut-offs and parameters of the theory constant. Similarly,
in gauge theories based on lattices, to compute expectation values of
Wilson loops in the continuum limit, one only allows rectangular
lattices and the refinement of these lattices is often tailored to the
Wilson loop in question. Indeed, some of these strategies seem so
`natural' that the restrictions involved often go unmentioned.

To remove the regulator in the quantum theory, we will proceed as
follows. First, we will fix a graph $\g$ and consider the subspace
$\cyl_{R(\g)}$ of $\cyl$ consisting of all states $\Psi_{\g'}$ which
are cylindrical with respect to a graph $\g'$ whose range coincides
with the range of $\g$. The regulated operators $\q_C$ and
$\hat{V}_R^{\B}$ leave this subspace invariant. Hence, we can
meaningfully focus just on $\cyl_{R(\g)}$ and specify how the
regulator is to be removed to obtain the operator $\hat{V}_R^\g$ on
$\cyl_{R(\g)}$ from $V_R^{\B}$.
For that we will use $\g$. However,  we will see that the operators will
stay unchanged if we choose another graph $\g'$ which has the same
range as $\g$.  Moreover, the operators will in fact preserve the
space $\cyl_\g$ of the cylindrical functions based on $\g$.  By
varying $\g$, we will thus obtain a family of operators on various
$\cyl_\g$. Finally, we will verify that they are compatible in the
appropriate sense \cite{7}, i.e., together constitute a well-defined
operator $\hat{V}_R$ on $\H$.

Let us then begin with the first step. Fix a graph $\g$ and focus on
$\cyl_{R(\g)}$. The allowed refinements of the partition of the region
$R$ will depend on the graph $\g$. More precisely, we will assume that
(for sufficiently small $\e$) the permissible partitions $\B$ satisfy
the following three conditions (see fig.2):
\begin{itemize}

\item[$(i)$] every vertex of the graph $\g$ (within $R$) is contained
in the {\it interior} of one of the cells, say $C$, and coincides with
the intersection point of the triplet of two-surfaces $S_1,S_2,S_3$
assigned to $C$ by the partition $\B$;
\item[$(ii)$] if a cell $C$ {\it does} contain a vertex, say $v$,
then $v$ is the unique isolated intersection point between the union
of the three two-surfaces $S_a$ associated to  $C$ and the range 
of $\g$; and,      
\item[$(iii)$] if a cell $C$ {\it does not} contain any of the
vertices, then the triplet of surfaces $S_a$ associated by $\B$ to $C$
intersects $\g$ at most at two points.  
 
\end{itemize}

\begin{figure}
\centerline{\epsfig{figure=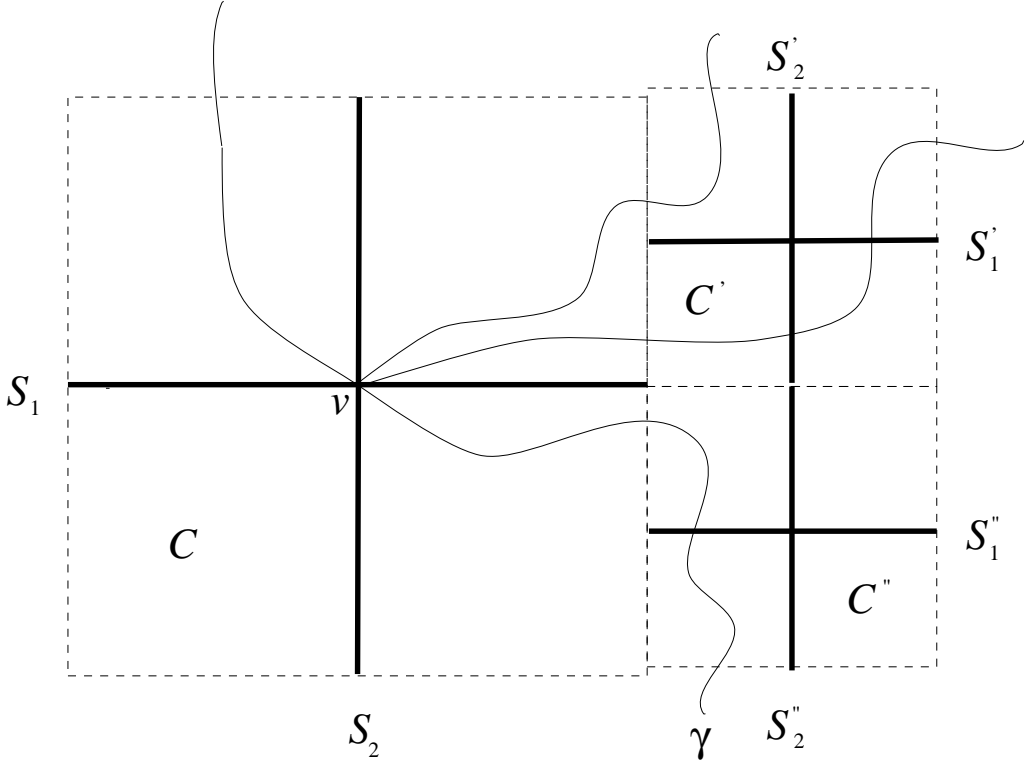,width=4in,angle=270}}
\caption{The figure illustrates the way cells $C, C', C''$ (the dashed
lines) and 2-surfaces $S_a, {S'}_a, {S''}_a$ (the bold faced lines) of
a partition $\B$ are adapted to a neighborhood of a vertex $v$ of a
graph $\g$ according to the conditions $i)-iii)$ above. For
simplicity, one dimension has been dropped.}
\label{fig2}
\end{figure}

These requirements are quite easy to meet. Given any partition $\B$
in which the vertices of the graph $\g$ do not lie on the walls of the
cells, the first condition can be met simply by choosing the surfaces
$S_a$ appropriately (within cells containing vertices).  Given a
partition satisfying the first condition, the second and third
conditions can be generically%
\footnote{The fact that the edges are not allowed to lie in the
surfaces $S_a$ in any cell does impose a mild restriction on the
permissible coordinate systems used to construct the partition $\B$.
However, this restriction is imposed only for simplicity of
presentation. Because of the averaging procedure of Section
\ref{sec4}, contributions from such non-generic partitions to the
final result are negligible.}
satisfied by a permissible refinement of that partition.  Furthermore,
once a refinement satisfying these two conditions is achieved,
subsequent refinements needed in the limiting procedure automatically
satisfy them. Nonetheless, these conditions do restrict the allowed
partitions. As we will see, they ensure that the limiting operator is
well-defined; if refinements are taken arbitrarily, in general the
limit fails to exist.

Let us now suppose that the partition $\B$ satisfies these conditions
and evaluate the action of the operator $\q_{C}$ on an element
$\Psi_{\g'}$ in $\cyl_{R(\g)}$. Note that because of the first
condition on the partition, a cell $C$ either contains one vertex of
$\g$ or no vertex at all.  If it does not contain a vertex, then by
condition $(iii)$ on the partition, due to anti-symmetrization forced
by $\eta_{abc}$, (\ref{hq1}) reduces to
\be\label{hq2}
\q_{C}\, \Psi_\g\ = \ 0 \ee
If the cell $C$ does contain a vertex, say $v$, then
(\ref{hq1}) and condition $(ii)$ on the partition implies:
\be\label{hq3} 
\q_{C}\,\Psi_\g =\ {1\over 48}\,\sum_{I,J,K}\, \e_{ijk}\e_{abc}\,
\k^a([e_I])\k^b([e_J])\k^c([e_K])J_{v,e_I}^iJ_{v,e_J}^j J_{v,e_K}^k \,
\Psi_\g\, , \ee 
where $I,J,K$ label the edges of $\g$ passing through the vertex $v$.
For simplicity, we will refer to $\k^a[e_I]$ as `vectors'.  The
possible `components' of these vectors are $0, \pm 1$ and depend only
on the octants defined by the two-surfaces $(S_1,S_2,S_3)$ which
contains the edge $e_I$.  To summarize, the conditions on permissible
partitions have streamlined the calculation by reducing (\ref{hq1}) to
(\ref{hq2}) and (\ref{hq3}).

Let us now focus on the non-trivial case in which $C$ contains a
vertex $v$.  The action of the operator $\q_C$ depends on the three
two-surfaces $S_a$ {\it only} through the properties of these surfaces
at $v$. Hence, it is unchanged as we refine the partition and shrink
the cell $C$ to $v$:
\be \label{hq42} 
\lim_{\e\rightarrow 0} \q_C\, \Psi_\g = 
{1\over 48}\, 
\sum_{I,J,K}\, \k([e_I], [e_J],[e_K])\,
\e_{ijk}\, J_{v,e_I}^iJ_{v,e_J}^j J_{v,e_K}^k \,
\Psi_\g\, ,
\ee
where the sum is over the edges passing through $v$ and where
\be\label{det2}
\k([e_I],[e_J],[e_K]) \,:=\, \e_{abc}\k^a([e_I])\k^b([e_J])
\k^c([e_K]).
\ee
(In particular, $\k([e_I],[e_J],[e_K])$ vanishes if any two of the
edges $e_I, e_J, e_K$ lie in the same or opposite octants.)  Using
the arguments used in the last sub-section, one can show that the
limiting operator is well-defined and self-adjoint on the Hilbert
space $H_\g$ obtained by Cauchy completing $\cyl_\g$.
Now, if given the space $\cyl_{R(\g)}$ we used in the above construction
a different graph $\g'$ whose range is the same as that of $\g$, the only
difference is that $\g'$ may have some extra bi-valent vertices
and/or some bi-valent vertices of $\g$ may be missing in the set
of vertices of $\g'$. However, for a bi-valent vertex, the operator
$\q_C$ vanishes identically. Therefore the resulting volume
operator derived in $\cyl_{R(\g)}$ using $\g'$ coincides with that for $\g$.

Unfortunately, however, this limiting operator (\ref{hq42}) carries a
memory of our choice of partitions through the term $\k([e_I], [e_J],
[e_K])$, i.e., on the background structure used in the regularization
procedure. Hence, although the limit of the operator
$\hat{V}_R^{(\B)}$ of (\ref{hv1}) is well-defined, it does not lead to
a viable candidate for the volume operator. However, since the
background dependence is of a rather simple type, one can eliminate it
by suitably `averaging' the regularized operator over relevant
background structures. We will carry out this averaging in the next
section.

To conclude this sub-section, let us note two properties of the
limiting operator (\ref{hq42}) which follow by inspection. {\it
Irrespective of the choice of permissible partitions used in the
regularization procedure}, we have:

\begin{itemize}

\item[$(1)$] $\k([e_I],[e_J],[e_K])$ depends {\it only}
on the germs of the edges; and,
\item[$(2)$] $\k([e_I],[e_J],[e_K])$ is totally anti-symmetric in its
three arguments. In particular therefore, if a graph has only bi-valent%
\footnote{If $\g$ has (at most) tri-valent vertices, simple algebra
shows that the action of the limiting operator on all {\it gauge
invariant} states in $\cyl_\g$ also vanishes \cite{16}. See Section
\ref{sec5}.}
vertices, the limit of the regularized operator $\q_C$ annihilates all
states in $\cyl_\g$.

\end{itemize}

We will see that these properties are trivially preserved by averaging
and are therefore shared by the final volume operator $\hat{V}_R$ we
will obtain in the next section.

\section{Averaging}
\label{sec4}

To remove the background dependence, we need to appropriately average
$\q_C$ over the relevant background structures, use the resulting
operator $\q_C^{\av}$ in place of $\q_C$ in (\ref{hv1},\ref{hq1}), and 
then take
the limit. Our discussion is divided in to four parts. In the first,
we spell out the basic strategy. In the second, we show that the
desired symmetries of the final volume operator and the consistency
requirement on the regularization procedure fix the form of the
averaged quantity $\k^\av ([e_I],[e_J],[e_K])$ ---and hence also of
the final volume operator--- uniquely up to a multiplicative
constant. In the third, we establish the existence of certain measures
that are needed for the averaging procedure. Finally, in the forth
part, we collect results of the Sections \ref{sec3} and
\ref{sec4.1} - \ref{sec4.3} to arrive at the desired volume operator.

\subsection{Basic Strategy}
\label{sec4.1}

The background dependence in the limit of the regularized volume
operator (\ref{hv1},\ref{hq42}) appears only through the factors
$\k([e_I],[e_J],[e_K])$ associated with cells containing a vertex of
$\g$. Therefore, let us first focus only on a single cell $C$
containing a vertex $v$. Note that $\k([e_I],[e_J],[e_K])$ depends
only on the relation between the three edges and the coordinate
octants {\it at} the vertex $v$.  Therefore, while there is an
infinite-dimensional freedom in the choice of the background
coordinates with which we began, the `relevant' freedom for averaging
turns out to be only finite dimensional. To see this, let us regard
two coordinate systems in $C$, centered at $v$, as equivalent if they
yield the same $\k([e_I],[e_J],[e_K])$ for all triplets of edges of
$\g$ passing through $v$.  Then, given two equivalence classes and a
coordinate system belonging to the first, one can generically
\footnote{We can focus on the generic case because the averaging
procedure involves integration and lets us ignore `sets of measure
zero'.}
find a system in the second which is related to the first by the
action of $GL^+(3)$, the group of orientation-preserving general
linear transformations at $v$. Furthermore, the diagonal subgroup
$\diag$ of $GL^+(3)$ merely re-scales the coordinates and hence leaves
each $\k^a[e_I]$ unchanged. Hence, to get rid of the background
dependence, it suffices to average $\k([e_I],[e_J],[e_K])$ only on the
{\it finite} dimensional coset-space $GL^+(3)/\diag$. Topologically,
this space ${\cal S}$ of `relevant' background structures can be
identified with an open subset of $S^2\times S^2\times S^2$ and
coordinatized by six angular coordinates, $\theta^A$ say, with
$A=1, ... ,6$.

Our task is to average the operator $\q_C$ over the space ${\cal S}$
in such a way that the resulting volume operator $\hat{V}_R$ is
well-defined and covariant, i.e., has no memory of the background
structures used in the regularization procedure. Fix a coordinate
system $x^a$ in an open region $U$ within $\S$ containing our region
$R$ and an adapted, permissible partition $\B$ of $R$ as in
Section \ref{sec3.3}. Fix a vertex $v$ of $\g$ and, as before, denote by $C$
the cell containing $v$. Given a $\theta^A \in {\cal S}$, we obtain a
second coordinate system $x^a(\theta)$, obtained from the first by the
action of an element of $GL^+(3)$ corresponding to $\theta^A$ (around
some fixed origin in $R$).  Using this coordinate system, we can again
construct a permissible partition $\B(\theta)$ of $R$. Denote by
$C(\theta)$ the cell in this partition containing the vertex
$v$. Replacing $C$ by $C(\theta)$ in Section \ref{sec3.3}, for each $\theta
\in {\cal S}$, we obtain an operator $\q_C^{\theta}$ on
$\cyl_\g^{(1)}$.  We wish to average these operators with respect to a
suitable probability measure on ${\cal S}$. Now, given {\it any}
normalized function $\mu(\theta)$, (i.e. satisfying $\int_{\cal S}\,
d^6\theta\, \mu(\theta) = 1$) the average $\q_C^{\rm av}$ of
$\q_C^\theta$ is given by:
\ba \label{4.1}
\q_C^{\av}\, \Psi_\g &=& \int_{\cal S}\, d^6\theta\, \mu(\theta)
\q_C^\theta  \Psi_\g \nonumber\\
&=& {1\over 48}\, \sum_{I,J,K}\, \k^{\rm av}([e_I], [e_J],[e_K])\,
\e_{ijk}\, J_{v,e_I}^iJ_{v,e_J}^j J_{v,e_K}^k \,
\Psi_\g\, , \ea
where 
\be\label{det3}
\k^{\av} ([e_I],[e_j],[e_K])\ =\ \int_{\cal S}d^6\theta\mu(\theta)
\k([e_I],[e_J],[e_K],\theta).
\ee
Thus, for any normalized measure $\mu$, the averaged operators
$\hat{q}_C^{\av}$ are well-defined. Using these in place of $\q_C$ in
(\ref{hv1}), one can construct the regularized volume operator
$\hat{V}_R^{\av}$. As in Section \ref{sec3}, it is straightforward to
remove the regulator. So far, for simplicity of presentation, we have
focussed on a single cell $C$ containing a vertex $v$. However, since
the coordinate systems $x^a(\theta)$ were not adapted to any specific 
cell, (\ref{4.1}) and (\ref{det3}) hold for all cells.

To summarize, the basic idea is simply to repeat the procedure of
section \ref{sec3} but now using {\it averaged} operators.  Physical
justification for this strategy comes from the fact that, as we saw in
Section \ref{sec3.1}, the averaging procedure does yield the correct
volume functional in the classical theory.

The key question then is: what measure should we use for averaging?
Unfortunately, the space ${\cal S}$ is non-compact and does not admit
a canonical normalized measure.  Can a suitable measure be perhaps
selected by examining the classical limit? The answer is in the
negative: As we saw in Section \ref{sec3.1}, although the averaging
procedure {\it is} applicable classically, the averaged regulated
volume tends to $V_R$ for {\it any} normalized $\mu(\theta)$. In the
quantum theory, on the other hand, this is not the case and we need to
find $\mu(\theta)$ for which the final volume operator is background
independent. In Section \ref{sec4.2} we will assume that such measures
exist and show that the requirement of covariance of the final volume
operator determines $\k^{\av}$ ---and hence also the final volume
operator--- uniquely up to a multiplicative constant. Existence of
measures of the required type will then be established in Section
\ref{sec4.3}.

\subsection{Uniqueness of $\k^{\av}$}
\label{sec4.2}

Let us suppose that there does exist a normalized measure $\mu$ on
${\cal S}$ such that the resulting volume operator $\hat{V}_R$
transforms covariantly under diffeomorphisms of $\S$. What can we say
about the corresponding $\k^{\av}([e_I],[e_J],[e_K])$? The purpose of
this sub-section is to show that, given any cell containing a vertex
$v$ of any graph, the quantity must be of the type:
\be \label{kappa2} 
\k^{\av}([e_I],[e_J],[e_K]) = \k_{(\mu)}\,\, \e (e_I, e_J, e_K)
\ee
for some (measure-dependent) constant $\k_{(\mu)}$, where $\e(e_I,
e_J, e_K)$ is the orientation function which equals $0$ if the tangent
directions%
\footnote{A germ $[e]$ at a point $x$ in $\S$ defines unique oriented 
tangent direction at $x$.}
to the three edges are linearly dependent at the vertex $v$, and $\pm
1$ if they are linearly independent and oriented positively or
negatively. (Recall that $\S$ is oriented.) Thus, the measure
dependence is contained in a single, overall multiplicative constant.

The idea is to use symmetries of $\k^{\av}([e_I],[e_J],[e_K])$ implied
by (\ref{det2}), (\ref{det3}) and the requirement that $\hat{V}_R$ be
diffeomorphism covariant, to constrain its form.  We saw at the end of 
the last sub-section that $\k ([e_I],[e_J], [e_K])$ has two properties
irrespective of the choice of a permissible partition. It is trivial
to verify that these properties are preserved by averaging. Thus, we
have:

\begin{itemize}

\item[$(1)$] $\k^\av([e_I],[e_J],[e_K])$ depends {\it only} on the
germs of the edges; and,
\item[$(2)$] $\k^\av([e_I],[e_J],[e_K])$ is totally anti-symmetric in
its three arguments.

\end{itemize}

Next, recall that a cylindrical function in $\cyl_\g$ is also
cylindrical with respect to a graph $\g'$ with $\g'\ge g$, i.e., such
that (the range of) $\g$ is contained in (the range of) $\g'$. Since
we want the volume operator $\hat{V}_R$ to be well-defined at the end,
its action on a state should not depend on whether we regard the state
as being cylindrical with respect to the first graph or the
second. This implies that

\begin{itemize}

\item[$(3)$] the function $(e_I, e_J, e_K) \, \rightarrow \,
\k^{\av}([e_I],[e_J],[e_K])$ defined by $\k^\av([e_I],[e_J],[e_K])$
depends only on the germs of the three edges and {\it not} on the
specific graph used in the computation.  Thus, the averaging procedure
simply provides a function from germs of (ordered) triplets of edges
intersecting at any point $x$ in $\S$ to reals.

\end{itemize}

Next, the assumed diffeomorphism covariance of the volume operator
implies that this function must have the following property:

\begin{itemize}
\item[$(4)$] Given any two triplets $(e_I, e_J, e_K)$ and
$(e'_I,e'_J, e'_K)$ of edges related by an orientation preserving
diffeomorphism of $\S$, 
\be 
\k^\av([e_I],[e_J],[e_K])\, = \, \k^\av([e'_I],[e'_J],[e'_K])\, .
\ee
\end{itemize}  

The next property follows immediately from (2)-(4). 
\begin{itemize}

\item[$(5)$] Let the triplet be such that the tangent directions they
define at $x$ are linearly independent. Then, 
\be \label{det4}
\k^\av([e_I],[e_J],[e_K]) = \k_{(\mu)}\,\, \epsilon(e_I, e_J, e_K)
\ee
since this is the only diffeomorphism invariant, totally
anti-symmetric function of germs of ordered triplets of edges
intersecting at $x$.

\end{itemize}

On the other hand, if the three tangent directions are linearly
dependent, other invariants exist, which depend on higher derivatives
of edges at the intersection point. Apriori, these are potential
candidates for $\k^\av([e_I],[e_J],[e_K])$. However, they can be ruled
out as follows. Consider two triplets with same orientation, $([e_1],
[e_2], [e_3])$ and $([e_1], [e_2], [e'_3])$ such that ${e'}_3$ is
tangent to $e_3$ at $x$. For a point $\theta\in {\cal S}$ such that
none of the corresponding two-surfaces $S_a$ passing through $v$ is
tangent to any of the germs $e_1,e_2,e_3$, the germs $e_3$ and
${e'}_3$ are on the same side of each of the two-surfaces.  Therefore,
almost everywhere on ${\cal S}$, we have $\k_{S_a}(e_3)\ =\
\k_{S_a}({e'}_3)$. Hence, whenever the three pairs of primed and
unprimed germs define the same oriented tangent directions at $x$, the
equality
\be  \ \ \ \k([e_I], [e_J], [e_K],\theta)\, = \, 
\k([e'_I], [e'_J], [e'_K],\theta) \ee 
holds almost everywhere on ${\cal S}$. In this case, integration with
respect to a $\mu(\theta) d^6\theta$ yields
\be
\k^\av([e_I],[e_J],[e_K])\, = \, \k^\av([e'_I],[e'_J], [e'_K]). \ee
To summarize, we have shown that
\begin{itemize}

\item[$(6)$] given a point $x$ in $\S$, $\k^\av([e_I],[e_J],[e_K])$
can depend {\it only} on the oriented tangent directions at $x$
defined by the three germs.

\end{itemize}

We can now use this property to evaluate $\k^\av([e_I],[e_J],[e_K])$
in the case when the tangent directions are linearly dependent. If two
of these directions coincide, anti-symmetry immediately implies that
$\k^\av([e_I],[e_J],[e_K])$ must vanish. Next, note that (\ref{kappa1})
implies that $\k_{S_a}[e_I] = -\k_{S_a}[-e_I]$. Hence,
$\k^\av([e_I],[e_J],[e_K])\,=\, -\k^\av([-e_I],[e_J],[e_K])$, which in
turn implies that $\k^\av([e_I],[e_J],[e_K])$ vanishes if any two
tangent directions are anti-parallel. Finally, consider the remaining
case in which the three tangent directions span a two-plane but are such
that no two of these directions are tangential to each other. Then,
modulo a possible diffeomorphism (to `straighten out the edges'),
there is a coordinate system in ${\cal S}$ in which (the point $x$
lies at the origin and) the three germs coincide with straight lines
$(1,0,0), (0, 1, 0)$ and $(1,1,0)$. Finally, there exists an
orientation preserving diffeomorphism --the rotation through $\pi$
along the third axis-- which carries $([e_1], [e_2], [e_3])$ to
$([e_2], [e_1], [e_3])$. Hence, by diffeomorphism invariance and
anti-symmetry, we conclude $\k^\av([e_I],[e_J],[e_K]) = 0$ in this
case as well. Thus, we have:
\begin{itemize}

\item[$(7)$] If the three tangent directions determined by the edges 
$([e_I],[e_J],[e_K])$ at $x$ are linearly dependent, then 
$\k^\av([e_I],[e_J],[e_K]) = 0$.

\end{itemize}

Properties (5) and (7) imply that {\it if} there exists a measure
$\mu$ on ${\cal S}$, averaging with respect to which provides a
well-defined diffeomorphism covariant volume operator $\hat{V}_R$,
then that averaging determines $\k^\av([e_I],[e_J],[e_K])$ uniquely,
up to a multiplicative constant $\k_{(\mu)}$. It is given by
(\ref{kappa2}).  Finally, properties (3) and (6) imply that the
constant $\k_{(\mu)}$ can depend only on the averaging function
$\mu(\theta)$ and not on the specific vertex or graph under
consideration.
\subsection{Existence of required measures}
\label{sec4.3} 

We now turn to the issue of existence: do there exist averaging
functions $\mu(\theta)$ which lead to volume operators $\hat{V_R}$
that transform covariantly under diffeomorphisms? We just saw in
Section \ref{sec4.2} that the necessary and sufficient condition for
$\hat{V_R}$ to be well-defined and covariant is that $\k^{av}$ of
(\ref{det3}) be given by (\ref{det4}). Therefore, we can rephrase the
question as follows: Given any vertex $v$ of any graph $\g$,
does there exist a function $\mu(\theta)$ on ${\cal S}$ such that,
for any triplet $e_I,e_J,e_K$ of edges of $\g$ at the vertex, 
$$ \int_{\cal S}d^6 \theta \mu(\theta) \k([e_I],[e_J],[e_K],\theta)\ 
=\k_o  \e([e_1],[e_2],[e_3]) \quad {\rm and} \quad 
\int_{\cal S} d^6\theta \mu(\theta) = 1 $$
for some constant $\k_o$ ? 

Now, since $\k([e_I], [e_J], [e_K])$ depends only on the oriented
tangent directions of the three edges and since the integrals in the
above equations represent the $L^2$ inner product on $({\cal S},
d^6\theta)$, (of $\mu$ with $\k$ and $1$ respectively) the required
$\mu(\theta)$ is guaranteed to exist provided the following statement
holds:
\smallskip
\noindent {\it For any finite set of germs $\{[e_1],...,[e_N]\}$ at a
point $v$, such that $e_I$ is not tangent to $e_J$ whenever $I\not=
J$, the functions $\k([e_I], [e_J], [e_K], \theta)$, $I<J<K$ and the 
constant function constitute a set of linearly independent 
functions on ${\cal S}$.}
\smallskip

Let us therefore establish this statement. Recall first that the set
of regulators $\cal S$ can be identified with the set of triplets of
oriented two-surfaces, $\theta \equiv (S_a)_{a=1,2,3}$, intersecting
at $v$. (These triplets are obtained by applying the orientation
preserving $GL(3)$ linear transformations on the surfaces $x^a =
const$, the action of $GL(3)$ being defined with respect to this
initial coordinate system.) Now, given germs $[e_1],...,[e_N]$ as in
the statement, suppose that there exist constants $a,\,a_{IJK}$,
$I,J,K=1,...N$ such that
\be\label{id?}  
a\ +\ \sum_{I<J<K} a_{IJK}\k([e_I], [e_J], [e_K],\theta) =\ 0, \ee 
for almost every $\theta=(S_1,S_2,S_3)\in {\cal S}$.  Pick a
two-surface $S_0$ which is tangent to $[e_1]$ but is not
to any of the other germs. Consider the points of ${\cal S}$
where $S_1$ is near $S_0$. Then, as we slightly vary $S_1$ around
$S_0$, the functions $\k_{S_1}(e_J)$ remain unchanged if $J\not=1$. On
the other hand, $\k_{S_1}(e_1)$ does change. We can find two-surfaces
$S_-$ and $S_+$ near $S_0$ such that
\be \k_{S_\pm}(e_I)\ =\ \pm 1.  \ee 
Therefore, plugging into (\ref{id?}) $\theta_-:=(S_-,S_2,S_3)$, and
next $\theta_+=(S_+,S_2,S_3)$ with any $S_2,S_3$, and subtracting, we
find 
\be  a_{1JK}\,\e_{1bc}\, \k_{S_b}(e_J)\k_{S_c}(e_K)\ =\ 0 \ee
for arbitrary $S_2,S_3$. Repeating that argument for $e_2$ and $e_3$,
say, we obtain
\be a_{123}\ =\ 0.  \ee 
Since this argument applies to any triple $I,J,K$, we conclude
\be a_{IJK}\ =\ 0\ =\ a.  \ee 
Thus, as asserted in the statement above, the constant function and
the functions $\k([e_I], [e_J], [e_K], \theta)$ are all linearly
independent on ${\cal S}$.

This establishes the existence of averaging functions of the required
type.

\subsection{Summary} 
\label{sec4.4}

Let us now collect the results obtained in Section \ref{sec3} and the
previous three subsections. In Section 3, we began by recasting the
volume function $V_R$ on the classical phase space such that the
resulting `regulated' form could be taken over to the quantum theory.
We then showed that if the regulator is removed with due care, the
limiting operator (\ref{hv1}, \ref{hq42}) is well-defined. However, it
carries a memory of the regulators used and therefore the resulting
volume operator fails to be covariant under diffeomorphisms. To remedy
this situation, in Section \ref{sec4} we introduced a procedure to
average over the `relevant' background structures. It turns out that
the requirement that the final operator $\hat{V}_R$ be well-defined
and diffeomorphism covariant is so strong that it determines the form
of the averaged operator except for an overall multiplicative constant
$\k_o$. Finally, this form does result from averaging with respect to
suitable measures, i.e., the averaging functions of the required type
do exist.

The final result is the following. Given a region $R$ in $\S$, we have
an operator $\hat{V}_R$, whose action on any cylindrical function
$\Psi_\g$ is given by
\ba\label{volgamma}
\hat{V}_R\,\Psi_\g\ &=&\ \k_o \sum_v \sqrt{|\q_v|}\, \Psi_\g \quad
{\rm where} \nonumber\\ 
\q_v\, \Psi_\g\ &=&\ {1\over 48}\e_{ijk}\sum_{e,e',e''}\e(e,e',e'')
J^i_{v,e}J^j_{v,e'}J^k_{v,e''}\, \Psi_\g \, .  
\ea
Here $v$ runs over the set of vertices of $\g$; $e,e',e''$ over the
set of edges of $\g$ passing through the vertex $v$ and $\e(e,e',e'')$
is the orientation function, which equals $0$ of the tangent
directions of the three edges are linearly dependent and $\pm 1$ if
they are linearly independent and oriented positively or negative with
respect to the fixed orientation on $\S$.  The constant $\k_o$ remains
undetermined. As written, because of the explicit reference to the
graph $\g$, the operator is defined in $\Cyl^{(3)}_\g$ (for any $\g$).
Next, it is quite easy to check that for any pair of graphs, the
corresponding operators agree on the intersection $\cyl^{(3)}_\g
\cap\cyl^{(3)}_{\g'}$. That is, the family of operators defined on various
$\cyl^{(3)}_\g$ satisfy the `cylindrical consistency' condition introduced
in \cite{7}). Therefore, (\ref{volgamma}) unambiguously defines an
operator in $\Cyl^{(3)}$.
 
We conclude with two remarks.

1) {\it Restoration of constants}: Let us restore the factors of $c$,
$G$, $\hbar$ and the Immirzi parameter $\gamma$ in the final
expression.  In the quantum representation labelled by $\gamma$, the
volume operator is given by:
\ba\label{volgamma'}
\hat{V}_R\,\Psi_\g\ &=&\ \k_o (\frac{8\pi G\hbar\gamma}{c^3})^{3\over 2} 
\sum_v \sqrt{|\q_v|}\, \Psi_\g \quad {\rm where} \nonumber\\ 
\q_v\, \Psi_\g\ &=&\ {1\over 48}\e_{ijk}\sum_{e,e',e''}\e(e,e',e'')
J^i_{v,e}J^j_{v,e'}J^k_{v,e''}\, \Psi_\g \, .  
\ea
In the remainder of the paper, however, we return to the conventions
$c=1$, $8\pi G =1$, $\hbar =1$ and $\gamma =1$.

2) {\it Extension to the smooth case}: As we remarked in footnote 2,
the analyticity of the graphs is not really essential for the results
presented above. We can begin with the vector space spanned by the set 
of cylindrical functions
$\cyl_\infty$ given by all the smooth graphs, replacing `analytic' by
`smooth' in the defintion of a graph, following \cite{LM}. The work of
Baez and Sawin \cite{5} provides a natural extension of the integral
$\int d\mu_o$ defined for elements of $\cyl_\infty$. The Cauchy
completion of this space then leads to a Hilbert space. In the
regularization of volume, the only potential problem with this
extension is that the domain of the operator $\E^i_S$ smeared over a
two-surface $S$ (which may or may not be analytic) fails to be dense
in the Hilbert space.  This problem occurs because a smooth graph can
intersect a given two-surface in an infinite number isolated points.
Therefore, to define the regulated operators (\ref{hv1},\ref{hq1}) we
need to start with a partition $\B$ that satisfies the conditions
$i)-iii)$ of Section \ref{sec3.3}. Then, given a graph $\g$ the number
of intersection between the two-surfaces $S_a$ defined by $\B$ and
$\g$ is finite by construction.  Hence, the two surface operators
$E^i_{S_a}$ are well defined in $\cyl_{R(\g)}$ and the entire
construction goes through.  The resulting volume is given by the same
formula (\ref{volgamma'}) and has all the properties discussed
below. One of them, the diffeomorphisms invariance, is even easier to
formulate because the action of the smooth diffeomorphisms is now well
defined on the whole Hilbert space. (However the arguments used 
should be modified: one has to use the spin-network decomposition
from the beginning.) Using analogous modifications, the
`external' regularization of volume presented in the Appendix can also
be extended to that smooth case.  

\section{Properties of the volume operator.}\label{sec5}
Although the volume operators $\hat{V}_R$ are not as well understood
as their area counterparts, their basic properties have been explored.
The purpose of this section is to summarize these.

\subsection{Preliminaries}
\label{sec5.1}  

1) As with area operators, the expression of $\hat{V}_R$ can be recast
in an `intrinsic' fashion that does not refer to any graphs.
\ba\label{vfinal} 
\hat{V}_R\ &=& \ \k_o\sum_{x\in R} \sqrt{|\q_x|},\quad {\rm where}
\nonumber\\ 
\q_x\ &=& {1\over 48} \sum_{[e],[e'],[e'']} \e_{ijk}\,
\e(e,e',e'')\, J^i_{x,e}J^j_{x,e'}J^k_{x,e''} \ea 
where each of $[e]$, $[e']$ and $[e'']$ runs through the set of germs
of one-dimensional sub-manifolds of $\Sigma$, bounded from one side by
$x$. As usual, acting on a cylindrical function $\Psi_\g$ the action
is non-trivial only if $x$ is a vertex of the graph and germs
$[e],[e'],[e'']$ overlap three edges of $\g$ intersecting at $x$.  In
the terminology of \cite{7} the operator $\q_x$ is given by a
cylindrically consistent family - labelled by all the graphs - of
essentially self-adjoint operators $\cyl_\g^{(3)}\rightarrow
\cyl^{(0)}_\g$. Hence, with domain $\cyl^{(3)}$, it is an essentially
self-adjoint operator on $\H$. Therefore the absolute value and square
root of this operator used in the first equality of (\ref{vfinal}) are
well defined and $\hat{V}_R$ is essentially self-adjoint.
 
2) In view of the above formulas it is meaningful to regard the
operator $\widehat{\sqrt{q}(x)}$, representing the square root of the
determinant of the metric as an operator-valued distribution:
\be
\widehat{\sqrt{q(x)}}\, \Psi_\g \, = \, {\k_o\over \sqrt{{48}}} 
\sum_v\, \delta^3(x,v)\, |\sum_{[e],[e'],[e'']}\, \e(e,e',e'')\, 
J^i_{x,e}J^j_{x,e'}J^k_{x,e''}|^{\frac{1}{2}}\, \Psi_\g \, .
\ee
By contrast, the determinant of the metric fails to exist even as an 
operator valued distribution.  This is completely analogous to the
situation for determinants of 2-metrics that was encountered in the
discussion of area operators.

3) If $R(x,\e)$ is a family of neighborhoods which shrink to $x$ as
$\e\rightarrow 0$, then given any $C^3$ cylindrical state $\Psi_\g$, 
the limit 
\be \lim_{\e\rightarrow 0}\hat{V}_{R(x,\e)}\, \Psi_\g \ee 
exists but in general is {\it not} zero. This property plays an
important role in the recent regularizations of various operators
that arise in quantum dynamics \cite{19,thomasmatter}.

\subsection{The gauge invariance and diffeomorphism covariance.} 
\label{sec5.2}

Both $\hat{V}_R$ and $\q_x$ are gauge invariant. Therefore, they
naturally restrict to the operators in the space of gauge invariant
cylindrical functions $\cyl^{(3)}(\Ab/\Gb)$. 

Since the edges of our graphs are analytic, analytic diffeomorphisms
on $\S$ have a well-defined action on $\cyl^{(3)}$. The measure
$\mu^o$ on $\Ab$ is invariant under this action. Hence the action of
these diffeomorphisms preserves the inner product on $\cyl^{(3)}$ and
extends to an unitary action of all of $\H$.  The operators
$\hat{V}_R$ and $\hat{q}_x$ transform covariantly under this action.

There is however a larger group, ${\rm Diff}$ of smooth
diffeomorphisms of $\Sigma$ that one can consider. Given $\phi\in{\rm
Diff}$, we obtain an operator on $\H$ whose domain%
\footnote{If we work in the Hilbert space of \cite{LM} given by all
the smooth graphs, then the domain of each diffeomorphisms is the whole
Hilbert space.} 
is the linear span of all cylindrical functions $\Psi_\g$ such that
$\g$ is mapped by $\phi$ to an analytically embedded graph. For
notational simplicity, we will denote these operators also by
$\phi$. Their action is given by
\be
\phi\, \Psi_{\g} (\ab)\, = \, \psi(\ab(\phi(e_1)), ... ,\ab(\phi(e_N))),
\ee
where $\psi$ is the function on $[SU(2)]^N$ that determines $\Psi_\g$
(see (\ref{cylin})). The volume operators transform covariantly with
respect to these diffeomorphisms. That is, 
\ba 
\phi \hat{V}_R \Psi_\g\ &=&\ \hat{V}_{\phi(R)}\, \phi \Psi_\g,
\nonumber\\ 
\phi \q_x \Psi_\g\ &=&\ \q_{\phi(x)}\, \phi \Psi_\g.  \ea 
In particular if $\phi$ preserves $R$, we have $[\phi,\hat{V}_R]\ =\
0,$ and, if it preserves $x$, then $[\phi,\q_x]\ = \ 0$. Note,
however, that the volume operators fail to be covariant with a similar
action of an arbitrary homeomorphism. In particular, recall that
$\hat{V}_R$ annihilates $\Psi_\g$ if the incident edges at every vertex
of $\g$ are co-planar. However, for every graph there is a
homeomorphism that can map it into a graph with the above property,
which contradicts the covariance. How does this situation compare with
that in the classical theory? Since densities fail to have a
meaningful transformation property under general homeomorphisms, image
of the volume element under homeomorphisms may not even exist!

If the region $R$ is all of $\Sigma$, the total volume operator
$\hat{V}_\Sigma$ is diffeomorphism {\it in}variant. Therefore it
induces the operator in the space of diffeomorphism invariant states
\cite{9,LM}).

\subsection{The spectrum}
\label{sec5.3}

We will first show that irrespective of the choice of the open region
$R$, the volume operators $\hat{V}_R$ have the same, discrete
spectrum.  Note first that, for every cylindrical function $\Psi_\g$
$\hat{V}_R \Psi_\g$ as well as $\q_x\Psi_\g$ are cylindrical over the
same graph $\g$. Let us therefore fix $\g$ and consider the
restrictions of (\ref{vfinal}) to $\Cyl_\g$ which define operators
thereon for every region $R\subset\Sigma$ and every point $x\in
\S$. In $\Cyl_\g$ each of the operators $\q_v$ is a finite sum with
constant coefficients of the products of triplets of the operators
$J^i_{v,e}$. It follows from the well known properties of operators
satisfying the angular momentum commutation relations (\ref{Jcom})
that the spectrum of $\q_v$ in the completion ${\cal H}_\g$ of
$\Cyl_\g$ is discrete. Therefore, so is the spectrum of the
restriction of $\hat{V}_R$ to ${\cal H}_\g$ for $\hat{V}_R$ is a
finite sum of the commuting square roots of $|\q_v|$s. If for each
$\g$ we denote the span of the eigen vectors in ${\cal H}_\g$ by ${\cal
E}_\g$ then $\cup_\g {\cal E}_\g$ is dense in $\H$.  Hence the full
spectrum of $\hat{V}_R$ is given by the union of the spectra on the
spaces $\H_\g$.

If we vary the graph $\g$, {\it a priori} it is possible that the
spectra might change in a continuous manner. However, the spectrum of
$\q_x$ depends only on the relative orientations of the triplets of
oriented tangent directions defined by the edges at $x$. The set of
possible characteristics is thus countable.  Therefore, the full
spectrum of $\q_x$ is a countable union of the countable set of
different spectra in each $\H\g$. The same argument shows that the
full spectrum of $\hat{V}_R$ is countable.

The fact that spectrum of $\hat{V}_R$ is independent of an open $R$ is
a simple consequence of the fact that the eigen values of $\q_x$
depend only on the characteristics of a graph in a arbitrarily small
neighborhood of $x$. This property is shared by the classical volume
function $V_R$ (although in that case the allowed values of the
function span the entire non-negative half-line for any open region
$R$).

It is also clear from the commutation relations (\ref{Jcom}) that
there is a basis such that each of the operators $J_{x,e}^i$ is of the
form $i$ times a skew-symmetric, {\it real} and block-diagonal matrix
of finite dimensional blocks corresponding to graphs. (This is a
spin-network basis \cite{10,11}; see also \cite{1} for an extension of
the definition of spin-networks from the space of gauge invariant
cylindrical functions to the space $\Cyl$ of all the cylindrical
functions.) Since $\q_x$ is constructed from the products of the
commuting $J_{x,e}^i$ operators, the same is true for
$\q_x$. Therefore, if a real number $\lambda$ is an eigen value of
$\q_x$ then so is $-\lambda$ and the corresponding eigen vectors are
related to each other by the complex conjugation of the coefficients.

Given a graph $\g$ and a vertex $x$, $\q_x$ commutes with each of the 
operators  $J^i_{v,e}$ provided $v\not=x$ as well as with the 
Gauss constraint operator 
$${\cal G}^i_x = \sum_{I=1}^N J^i_{x,e_I},$$
where $I$ in the sum labels the edges at $x$. {}From  these, one can
construct the following set of commuting operators. For each vertex
$v\not=x$, number the edges at $v$ in an arbitrary manner, say 
are $e_1,...,e_N$, and for every $k\le N$ defines the following
operator
$$
(\sum_{I=1}^k J^i_{v,e_{I}})\, (\sum_{I=1}^k J^i_{v,e_{I}})\ =:\  
(\sum_{I=1}^k J_{v,e_{I}})^2, $$ 
These operators, together with $\G^{i_v}_v$ for every vertex $v$ of
$\g$ and fixed $i_v$ for each $v$, form the required commuting
set. Subspaces preserved by $\q_x$ can be therefore labelled by the
eigen values of these operators. Let us denote by ${\cal
T}_{x,l,j_1,...,j_n,M}$ the subspace corresponding to the eigen value
$l(l+1)$ of ${\cal G}^i_x {\cal G}^i_x$, the eigen values $j_I(j_I+1)$
of $J^i_{x,e_{I}}J^i_{x,e_{I}}$, $I$ labeling the edges at $x$ and to
$M$ labeling the eigen values of the remaining operators.

Whenever ${\cal T}_{x,l,j_1,...,j_n,M}$ is one-dimensional it is
necessarily an eigen-direction of $\q_x$. Furthermore, the
corresponding eigenvalue must be zero.  In practice, this is a
powerful argument to find the kernel of $\q_x$ and was used by Loll
\cite{16} to show that the volume operator must annihilate all gauge
invariant ($l=0$) cylindrical functions $\Psi_\g$ if all vertices of
$\g$ are three or lower valent (i.e., if the number of edges meeting
at any vertex is less than or equal to three).  If ${\cal
T}_{x,l,j_1,...,j_n,M}$ happens to be $2$-dimensional, then the
absolute value operator $|\q_x|$ relevant for the volume is
automatically diagonal therein. Such a space, namely ${\cal
T}_{x,{1\over2},j_1,j_2,j_3,M}$, emerges in the evaluation of the
explicit formula for Thiemann's Hamiltonian operator acting on three
valent spin-networks \cite{BDPR}. In the general case, ${\cal
T}_{x,l,j_1,...,j_n,M}$ is finite dimensional which still reduces the
eigen problem to diagonalization of finite dimensional matrices.

The operator $\q_x$ consists of terms of the form 
\be
\e_{ijk}J^i_{e_I}J^j_{e_J}J^k_{e_K}\ =:\ \q_{IJK}.  \ee 
An intriguing property of that expression is that it can be written as
\be \ \hat{q}_{IJK}\ =\ {1\over 4i} [(J_{e_J}+J_{e_K})^2 ,
(J_{e_I}+J_{e_J})^2].  \ee It was observed and used by Thiemann to
analyze the matrix of $\q_x$ in a four valent case
\cite{Thomas2}. That property is also related to the `apparently
anomalous' terms in the commutator of two area operators \cite{acz}.
Several examples of the eigenvalues and eigenvectors and other special
cases were studied in \cite{Loll,18,DePietri}.  Although the volume
operator studied in \cite{18} differs from ours, and that of
\cite{Loll} is derived within the lattice framework, in the 4 (or
lower) valent cases there is a simple relation between all these
operators. In particular, the lattice operator coincides with the
above operator, restricted to the cylindrical functions given by the
loops contained in a cubic lattice \cite{15}.

Finally, the spectrum of the volume is discrete in the sense it has a
countable number of elements. The complete spectrum is not known
explicitly (in contrast to the situation with area operators). Indeed,
we do not even know how `densely the eigenvalues are packed' on the
real line, or even what the smallest non-zero eigenvalue is.

\subsection{Example: A gauge invariant 4-valent vertex.} 
\label{sec5.4}

Since vertices at which three or less edges meet are `trivial' as far
as the operators $\q_x$ and $\hat{V}_R$ are concerned, the simplest
non-trivial case is that of a gauge invariant 4-valent
vertex. Therefore, to get a feel for the action of these operators,
let us therefore discuss this case in some detail.

Suppose $x$ is a four valent vertex of a graph $\g$ and consider
$\q_x$ acting in the corresponding subspace of gauge invariant
elements of $\Cyl^{(3)}_\g$.  Denote by $e_I$, $I=1,2,3,4$ the edges
intersecting at a vertex $x$. Gauge invariance of states in this subspace
is equivalent to the statement that Gauss constraint vanishes identically
on this subspace. Thus, at the vertex $x$ we have:
\be 
J^i_{x,e_1 }\ +\ J^i_{x,e_2}\ +\ J^i_{x,e_3}\ +\ J^i_{x,e_4}\ 
=\ 0.  \ee 
We use this equation to eliminate $J^i_{x,e_4}$ in favor of the
remaining $J$s. Then, we have
\ba
\q_x &=& \ {\k_o\over 8}\,\e_{ijk}\, \Big(\e(e_1,e_2,e_3)J^i_{x,e_1}
J^j_{x,e_2} J^k_{x,e_3} \, + \,\e(e_4,e_2,e_3)J^i_{x,e_4}
J^j_{x,e_2} J^k_{x,e_3} \nonumber\\
&+& \e(e_1,e_4,e_3)J^i_{x,e_1} J^j_{x,e_4} J^k_{x,e_3}\, +\,
\e(e_1,e_2,e_4)J^i_{x,e_1} J^j_{x,e_2} J^k_{x,e_4}\Big).  
\ea
Using the Gauss constraint and the observation that 
\be\label{id}
\e_{ijk}\, J^i_{x,e_1 }J^j_{x,e_2}(J^k_{x,e_1 }\ 
+\ J^k_{x,e_2})\ =\ 0
\ee
we find
\ba\label{4valent}
\q_x\ &=&\ \frac{\k_o}{8}\, \k(e_1,e_2,e_3,e_4)\, \e_{ijk}
J^i_{x,e_1 }J^j_{x,e_2} J^k_{x,e_3},\,\, {\rm where,} \nonumber\\ 
\k(e_1,e_2,e_3,e_4)\ &=& \ \e(e_1,e_2,e_3)\ -\ \e(e_1,e_2,e_4)\ -\
\e(e_1,e_4,e_3)\ -\ \e(e_4,e_2,e_3).
\ea
Thus, modulo the geometric factor $\k (e_1, e_2, e_3, e_4)$, the
action is the same as that at a three-valent, but non-gauge invariant,
vertex between the edges $e_1$, $e_2$ and $e_3$.
\footnote{It follows from the same arguments that, in a 4-valent case,
the Rovelli-Smolin counterpart (\ref{6.24}) of our $\q_x$ operator
(\ref{4valent}) coincides with $|\q_x|$ of (\ref{4valent}) with
$\k(e_1,e_2,e_3,e_4)=4$.}
(In a three-valent gauge invariant case, one can further express
$J^i_{x,e_3}$ as $J^i_{x,e_3} = - J^i_{x,e_1} - J^i_{x,e_2}$. Then
(\ref{id}) implies that $\q_x$ vanishes on this subspace.)

The value of the diffeomorphism invariant factor $\k$ depends on the
characteristics of the intersection between the ordered edges at $x$.
Recall from Section \ref{sec4.2} that it depends only on the oriented
directions $\de_I$, $I=1,...4$ of the vectors defined by the edges at
$x$.  If the intersection is planar (that is, the four edges are
tangent to a two-plane), then $\q_x$ is identically zero. Otherwise, we
can always find coordinates, such that after a possible renumbering,
\be \de_1=(1,0,0), \,\de_2=(0,1,0),\, \de_3=(0,0,1).  \ee 
Then, the intersection character is determined by $\de_4$.  One can see
that, every case (modulo renumbering) is diffeomorphic to one of the
cases given by the following possible values of $\de_4$:
\be
\de_4= (1,1,1),\, (-1,-1,-1),\,(1,1,0),\, (-1,-1,0),\, \de_3,\, -\de_3. 
\ee
The corresponding values of $\k(\e_1,e_2,e_3,e_4)$ are
\be
\k(e_1,e_2,e_3,e_4)\ =\ -2,\, 4,\, -1,\, 3,\, 0,\, 2.   
\ee
Finally, the dimension of the corresponding invariant subspace ${\cal
T}_{x0j_1,...,j_4,,M}$ is given by the number of different numbers $j$
each of which satisfies simultaneously: 
\be |j_1-j_2|\ \le j \le |j_1+j_2|, \ |j_3-j_4|\ \le j \le |j_3+j_4|, 
\ee 
and such that both $j-|j_1-j_2|$ and $j- |j_3-j_4|$ are integers.

\section{Discussion}\label{sec6}

In the main body of the paper we presented a regularization scheme to
obtain volume operators $\hat{V}_R$ associated with open regions $R$
of the `spatial' 3-manifolds $\S$ and discussed a few properties of
these operators. As in any `quantization', the idea is to first
express the classical observable of interest (in our case, functions
$V_R$ on the classical phase space) in terms of `elementary' variables
(in our case, two-dimensionally  smeared triads) which have
unambiguous quantum analogs, then promote this `regulated' classical
expression to quantum theory and finally remove the regulators.  There
is considerable freedom in the first step and we chose an `internal'
regularization in which the volume of an elementary cell is expressed
in terms of triads smeared over three two-surfaces passing through the
{\it interior} of the cell. It was relatively straightforward to carry
over the classical expression to the quantum theory and remove the
regulators. However, it turned out that the the resulting operator
carries a memory of the background structures used in the
regularization procedure and fails to transform covariantly under
diffeomorphisms of $\S$. To rectify this situation, we first averaged
the regulated expressions over the `relevant' background structures
and then removed the regulators. The resulting operator $\hat{V_R}$ is
uniquely defined up to an overall ($R$-independent) constant $\k_o$.
It is a densely defined, positive, self-adjoint operator and
transforms covariantly under the action of $\S$-diffeomorphisms. Its
spectrum is purely discrete.

The central part of the paper is contained in Sections \ref{sec3} and
\ref{sec4} which discuss the intricacies of the regularization
procedure. The overall philosophy here is the same as that used in
other quantum field theories and the subtleties involved in the
continuum limit are of the same nature as those encountered there.  In
our case, a key simplification occurs because (a large class of)
operators on the kinematical Hilbert space $\H$ can be naturally
considered as a consistent family of operators on partial Hilbert
spaces $\H_\g$ associated with graphs $\g$ \cite{7}.. In the present
case, in particular, we could focus on the restrictions
$\hat{V}_R^\g$ of the desired operator $\hat{V}_R$ to $\H_\g$. In the
regularization procedure, therefore, we could refer to the graph
$\g$. The key test of the procedure comes from consistency
requirement: the operators $\hat{V}_R^\g$ so regulated could have failed
to be consistent. This did not happen%
\footnote{In the initial stages, there was some concern because the
regularization procedure seemed to be `state-dependent'. Regarding
$\hat{V}_R$ as a consistent family of operators $\hat{V}_R^\g$
clarifies the situation: it is as natural to use $\g$ to regulate an
operator on $\H_\g$ as it is to use the Fock space structure to normal
order operators in Minkowskian quantum field theories.};
our family turned out to be consistent and therefore defines a
self-adjoint operator $\hat{V}_R$ on $\H$.

The regularization procedure could, however, be improved in some
respects.  First, the regulated expressions (\ref{q}, \ref{apprvol})
are not gauge invariant. However, a cosmetic change can rectify this
situation without affecting our arguments or the final result: it
suffices to replace the simple integration of $E$ in (\ref{q}) over a
two-surface with integration combined with the parallel transport
along some fixed paths ending at fixed point. This additional step
does not affect the result which is already gauge invariant. It would
be more difficult to make the regularization procedure manifestly
diffeomorphism covariant. However, this is largely an aesthetic issue
since our final result does enjoy diffeomorphism covariance. Next, at
least at first sight, the ambiguity of a multiplicative constant
$\k_o$ appears to be an undesirable feature. Recall however from
Section \ref{sec4.4} that there is another ambiguity at the
kinematical level, first pointed out by Immirzi \cite{immirzi} using
earlier work of Barbero \cite{barbero}. This arises from the existence
of a canonical transformation which fails to be unitarily
implementable and leads to a one parameter family of unitarily
inequivalent representations of the holonomy-triad operators. For
volume operators, the net effect is that there is an ambiguity of a
multiplicative constant in the spectrum which can not be eliminated
without additional input. Therefore, from a `practical' viewpoint, the
freedom in the choice of $\k_o$ does not worsen the situation.

The Appendix discusses another regularization scheme, based on
constructions given by Rovelli, Smolin and De Pietri \cite{12,13,18}
in the loop representation. This may be called an `external'
regularization because the starting point is an expression of the
volume of an elementary cell in terms of triads smeared on the {\it
boundary} of the cell. In the early discussions, it was believed that
this regularization is free of the subtleties we encountered in
Section \ref{sec3.2} while taking the continuum limit in the quantum
theory. However, a careful treatment of the limit shows that this is
not the case; the assumptions needed to ensure that a well-defined
limit exists are completely analogous to those introduced in Section
\ref{sec3.2}.  The final result does differ from that presented in the
main body of the paper. Because of the `external' regularization,
these regulated ---and hence also the final--- operators do not have
any information about the details of the tangent vectors to the edges
at vertices.  Hence they transform covariantly not only under
diffeomorphisms but also homeomorphisms on $\S$ in the sense spelled
out in Section \ref{sec5.3}. However, in simple situations such as
those considered in simplicial treatments \cite{barbieri}, the two
operators coincide (apart from a constant \cite{15}). Finally,
although it is often not explicitly stated, the ambiguity of an
overall constant exists also in the `external' regularization. There,
it is buried in the in one's choice of `rectangular' cells. If one
changes to tetrahedral cells, for example, the overall constant in
front of that volume operator would change. Indeed, the situation is
parallel to that in the `internal' approach of the main text. There,
the $\k_o$-ambiguity arose because, while the limit of the classical,
averaged expression is insensitive to the choice of the measure used
in the averaging, the limit of the quantum operator is unique only up
to an overall constant. Similarly, in the `external' approach, while
the limit of the classical, regulated expression is insensitive to the
details of the geometry of elementary cells, the corresponding quantum
volume operators can differ by an overall constant.

To conclude, we wish to emphasize that the existence of two distinct
operators is a reflection only of quantization ambiguities. In both
approaches, one begins by re-expressing the classical volume function
$V_R$ in terms of two-dimensionally smeared triads and takes these
`regulated' classical expressions over to quantum theory. The
difference lies in the choice of the `regulated' classical
expressions. In the classical theory, they both lead to the same
function $V_R$ when the regulators are removed. In the quantum theory,
there is a subtle difference in the corresponding operators
$\hat{V}_R$.

\bigskip

{\bf Acknowledgements.} We thank Roberto De Pietri for numerous
discussions about the volume operator, and for showing to us the loop
assignment which makes the `external' regularization convergent. One
of us (AA) was supported in part by the NSF grant PHY95-14240 and by
the Eberly research funds of Pen State. The other (JL) was supported
in part by the Alexander von Humboldt-Stiftung (AvH), the Polish
Committee on Scientific Research (KBN, grant no. 2 P03B 017 12) and
the Foundation for Polish-German cooperation.

\bigskip \bigskip

\appendix\label{sec7}
\centerline{\bf APPENDIX: `EXTERNAL' or `EXTRINSIC' REGULARIZATION}
\bigskip

In this Appendix, we will provide a detailed derivation of the
analytic formula of the volume operator (reported in \cite{15}) which
results from an `external' regularization scheme.  Key ideas behind
this regularization are due to Rovelli and Smolin \cite{17}. However,
our treatment differs from theirs in some ways. First, we work in the
connection ---rather than loop--- representation. Second, as in
Section \ref{sec3}, the removal of regulators is a subtle procedure
also in this `external' regularization and the limiting operator can
carry a memory of the coordinate system used or may not even exist
unless the permissible partitions are restricted suitably.  These
subtleties, discussed in Subsections \ref{secA.3} and \ref{secA.4}
below, were overlooked in \cite{17}.  Finally, as is the case for the
`internal' regularization discussed in the main text, our `external'
regularization is consistent with the treatment of the Hamiltonian
constraint in the literature \cite{19,LM}. Let us discuss this point
in some detail. Recall first that the domain of this constraint
operator is the diffeomorphism invariant subspace of the algebraic
dual to the Hilbert space $\H$ \cite{19}, or an appropriate extension
thereof \cite{LM}.  The topology one uses to take the limit while
removing the regulator is defined as follows. If $\hat{\cal
O}^\epsilon$ denotes the regulated operator on $\H$ (where $\epsilon$
symbolizes the regulator) and $\bra S|$ a dual state, then the limit
$\hat{\cal O}$ as $\epsilon \rightarrow 0$ is defined via its action
$\bra \hat{\cal O} S|$ on $\bra S|$ by the following formula
\be\label{top} \bra \hat{\cal O}S| \Psi \ket := \
\lim_{\epsilon\rightarrow 0}\, \bra S| \hat{\cal O}^\epsilon
\Psi\ket \ee
for every $\Psi\in\H$ in the domain of $\bra S\hat{\cal O}|$. In the
case of the volume operator now under consideration, it appears that
the topology used implicitly in \cite{17} is inequivalent to
(\ref{top}). (See \cite{crjl} for the discussion of that topology.)
Indeed, if in the derivation presented in \cite{17}, or in the one
proposed in \cite{Thomas2}, we take the final limits using the
topology corresponding to (\ref{top}), the final result would be
different from that given in \cite{17,18,Thomas2} (see footnote
13). In our treatment, as in the main text, we will remove the
regulator using the Hilbert space topology and obtain a densely
defined operator $\hat{V}_R$ on $\H$. The dual action of this operator
on the dual space coincides with the action defined by (\ref{top}).
Thus, while our final answer is equivalent to that of \cite{17}, our
regularization provides a general procedure which is `uniformly'
applicable to the volume operator as well as the Hamiltonian
constraint. Not only is this aesthetically pleasing but quite
essential if, for example, one wishes to incorporate the cosmological
constant in the Hamiltonian constraint.

As before, one begins with a partition $\C$ of the open region $R$ in
to cells and but associates to each cell $C$ extra paths to parallel
transport the triads $E$ to a fixed point. Let us choose an arbitrary
partition $\C$ of $R$ into cells.  In particular, $\C$ may be the
Rovelli-Smolin partition, obtained by fixing coordinates $x^a$
covering $R$ and introducing a family of two-surfaces $x^a\ =
n\epsilon$, where $a=1,2,3$ and $n$ ranges the integers, the cells
being given by the coordinate cubes. We will see however that to
remove the regulator in the quantum theory one can not just shrink the
size of these cells; unless the partition is restricted in a manner
similar to that of Section \ref{sec3}, contributions from some cells
may diverge in the limit $\e\goesto 0$.  Furthermore, even for such
restricted partitions, to ensure that the limiting operator is
well-defined, one has to choose paths in a specific manner, following
a prescription given by De Pietri \cite{12}. Roughly speaking, these
restrictions specialize the permissible partitions to the `same
extent' that the restrictions of Section \ref{sec3} do in the
`internal' regularization scheme.

\subsection{The Rovelli-Smolin strategy} \label{secA.1}

In this sub-section will outline the overall strategy of Rovelli and
Smolin using, however, the connection representation.  

Let $R$ be an open subset of $\S$ as before. In general, one may not
be able to cover $R$ with a single chart. However, the argument
presented in the beginning of Section \ref{sec3} will continue to
apply.  We can therefore restrict our attention to a topologically
trivial region on which a global chart does exist.

As in Section \ref{sec3.1}, let us then cover $R$ with a global
coordinate system and let $\C$ be the partition of $R$ into cubic
cells.  By $\pC$ we will mean the differential part of the topological
boundary of the cell $C\in \C$. For every point $z\in \pC$ which
belongs to (the interior of) a face of $C$ we wish to smear $E$ over
$\pC$ in a gauge covariant manner. For this purpose, to every
$z'\in\pC$ let us assign a path $p_{z'x_0}$ connecting it with a fixed
point $x_0$ in $C$. For later convenience, we will assume that the
path assignment is such that, as we shrink the size of the cell, for
every path $p_{zx_0}$ we have:
\be \label{pr} U_{zx_0}(A)\ \goesto\ 1\in SU(2) \ee
{\it uniformly} with respect to $z\in\pC$, for all the $C$s in $\C$.
{\it A cell partition $\C$ with such a path assignment will be denoted
by} $\P$. Finally, let us introduce a non-negative two point field
$f(z,z')$ on $\pC$, which is a density of weight one in $z$ and a
scalar in $z'$, such that
\be \int_{\pC}\, f(z(r,s), z') drds = 1\, . \ee

With this machinery at hand, the covariantly regularized classical
momentum $E_{\pC, f}(z)$ is defined as
\be\label{Efinv} 
E_{\pC, f}(z)\ =\ {1\over 2} \int_{\pC} f(z,z')(U_{p_{z'x_0}})^{-1}E^c(z')
U_{p_{z'x_0}}\, \eta_{abc}\,\, d{z'}^a\wedge d{z'}^b.  \ee
The Rovelli-Smolin regularized, squared volume of $C$ is given by%
\footnote{The use of the Rovelli-Smolin loop variables introduces
some more freedom in the choice of the extra paths, but that won't be
relevant for our arguments. Our aim is to make comparison with
\cite{17} easier by writing various formulas in the notation used
there.}
\be\label{RSdet'} 
q_{C}(E)\ :=\ {1\over 12}\int_{(\pC)^3}\,  
|\Tr (E_{\pC, f}(z_1) E_{\pC, f}(z_2) E_{\pC, f}(z_3))|\,
d^2z_1 d^2z_2d^2z_3\, , \ee 
where by $d^2z$ we mean the parametrization dependent area element on
$\pC$ given by any parametrization thereof.  Since the gauge
transformation law is $E_{\pC, f}(z)\mapsto g^{-1}(x_0) E_{\pC,
f}(z)g(x_0)$, the trace is gauge invariant. Finally, set
\be\label{RSvol} V_{R}^{\P}(A,E)\ :=\ \sum_{C\in\C}
\sqrt{q_{C}(A,E)}.  \ee
This is the Rovelli-Smolin regulated volume functional on the
classical phase space. As we shrink cells, we have: $V_{\P}(A,E)
\goesto V_R(E)$. The idea is again to promote the regulated expression
$V_R^{\P}$ to the quantum theory and then remove the regulators.

\subsection{Regulated operators} \label{secA.2}

As in the main paper, we will first focus on a graph $\g$ and study
the action of the regulated operators on the Hilbert space
$\H_\g$. The overall procedure is parallel to the one adopted in
Section \ref{sec3}: the main idea is to replace every $E_{\pC, f}(z)$
in the functional $q_{C}(A,E)$ by the operator $\E_{\pC, f}$.

Let us then begin with an arbitrary partition $\P$ and consider
cylindrical functions $\cyl_\g(\Ab)$ compatible with a graph $\g$.  We
will only assume that the partition is generic with respect to
$\g$. The operator $\Tr (\E_{\pC, f}(z_1) \E_{\pC, f}(z_2) \E_{\pC,
f}(z_3))$ acts on elements of $\cyl_\g$ through the operators
$J^i_{w_I,p_I}$ of (\ref{Jop}), where $w_I$ are the intersection
points of the edges $e_I$ of $\g$ with $\pC$,
\be J_{w_I}^i\ :=\ {1\over 2}\sum_{p_I\ {\rm at}\ w_I}
\k_C(p_I)\, J_{w_I,p_I}^i \ee 
where $p_I$ runs over the set of segments of edges of $\g$
intersecting $w_I\in \pC$. Let $p_{x_0z}$ be the paths in the partition
$\P$ and set
\be U(z,z')\ :=\ U_{p_{zx_0}}U_{p_{x_0,z'}}\, . \ee 
Then, 
\ba\label{|.|} 
\Tr (\E_{\pC, f}(z_1) \E_{\pC, f}(z_2) \E_{\pC, f}(z_3))\psi\ =\
\sum_{w_I,w_J,w_K\in \pC}\,
f(z_1,w_I)f(z_2,w_J)f(z_3,w_K)\nonumber\\
\times\, \Tr\Big(\tau_i U(w_I,w_J)(A) \tau_j U(w_J,w_K)(A) \tau_k 
U(w_K,w_I)\Big) J_{w_K}^k J_{w_J}^{j}J_{w_I}^{i},
\ea
where the points $w_I, w_J, w_J$ on $\pC$ contribute only if they
simultaneously coincide with the isolated intersection points of $\g$
with $\pC$.

Let us denote the right side of of (\ref{|.|}) by $\q_{z_1z_2z_3}$.
It we could show that it is self-adjoint, then the quantum version
$\int d^2z_1 d^2z_2 d^2z_3|\q_{z_1z_2z_3}|$ of (\ref{RSdet'}) could be
well-defined. We will carry out these steps in detail later.  It turns
out that even when these steps are completed, there are obstacles in
removing the regulators. It is convenient to first discuss how they
arise and how they can be overcome by appropriately restricting the
limiting procedure.

\subsection{Ensuring the convergence I: Restrictions on cells} 
\label{secA.3}

Given $\Psi_\g\in \cyl$, it turns out that the shrinking of cells in
$\P$ is not sufficient for the vector $V_R^{\P}\Psi_\g$ to
converge. Let us isolate the potential problems. Let $\g$ be a graph
compatible with $\Psi_\g$. So far, we have only assumed that the
partition $\P$ is generic, i.e., that no vertex of $\g$ intersects any
$\pC$.  (This property can be ensured by assuming that $\P$ is shrunk
in such a way that no segment of any cell remains fixed, the
assumption being graph $\g$ independent.) Now, if we assume that the
partition has been sufficiently refined, say by the rescaling the
coordinates, the cells of $\P$ can be classified in the following
way:\\ {\it Type (i)}: $\pC$ surrounds a vertex of $\g$ and intersects
every edge which meets that vertex exactly once;\\ {\it Type (iia)}:
$\pC$ does not surround any vertex of $\g$ and intersects at most one
edge of $\g$ in at most two points;\\ {\it Type (iib)}: $\pC$ does not
surround any vertex of $\g$ and intersects more than one edge of $\g$.

As the discussion of \cite{17} and Section \ref{sec3} suggests, it is
only the type {\it (i)} cells that should contribute non-trivially to
the final result. Indeed, for a cell of type {\it (iia)},
\be \q_{\P} \Psi_\g\ =\ 0 \ee 
because, as we will see below, only $w_1\not=w_2\not=w_3\not=w_1$
contribute to (\ref{|.|}); cells of this type can be ignored.  What
may be easily overlooked is that one can not get rid of cells of type
{\it (iib)} so simply; these contributions can be coordinate dependent
and may even diverge.  To see this, suppose we just use the partition
$\P_L$ given by a cubic lattice of the size $L$ defined by the fixed
coordinates system on $R$. Consider a case when two edges $e_1$ and
$e_2$ intersecting at $v$ cross the same face of a cell $C_v$
containing $v$ (see Fig.3). The other cell $C$ which shares that face
is crossed by the edges $e_1$ and $e_2$.  This produces a non-zero
term $\q_{C}\Psi_\g$. One may, of course, choose another coordinate
system such that $e_1$ and $e_2$ intersect disjoint faces of ${C'}_v$
and do not cross the same adjacent cell.  But this simply serves to bring
out the ambiguity caused by the {\it coordinates dependence}.
Shrinking of $\P$ does not eliminate that ambiguity which is sensitive
to rotations of a given $C_v$ rather than to its size.

\begin{figure}
\centerline{\epsfig{figure=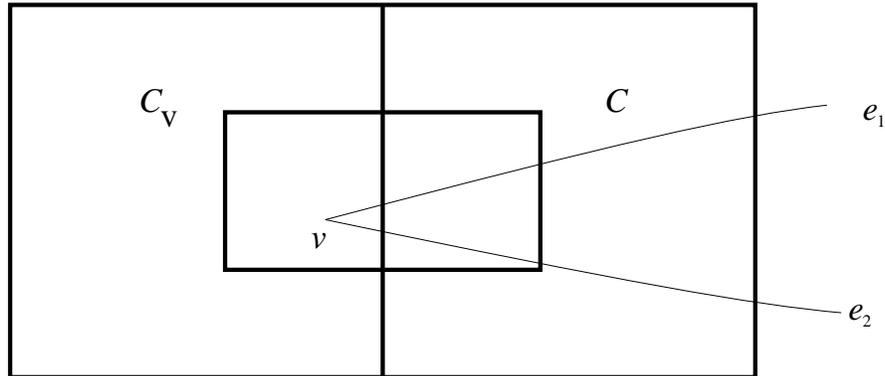,width=2in,angle=270}}
\caption{The size change of $C_v$ and $C$ does not change the fact, that
$e_1,e_2$ intersect the neighboring cell $C$.} 
\label{fig3}
\end{figure}

For similar reasons, the limiting procedure produces divergences when
$e_1$ and $e_2$ are tangent to each other at $v$. In this case,
generically, as one shrinks the maximal size of cells, the number of
cells crossed simultaneously by $e_1$ and $e_2$ tends to {\it infinity}. One
can check that the consequence is that the norm $\|\hat{V}_\C\Psi_\g\|$
diverges in the limit.

To summarize, {\it to ensure convergence, as in Section} \ref{sec3},
{\it one has to put restrictions on how the limit is taken using the
graph $\g$.} The choice we will make ---which seems to be the only
natural one available--- is to restrict ourselves to partitions
containing cells only of the types $(i)$ and $(iia)$.  This
restriction is very similar to that used in Section \ref{sec3}.

\subsection{Ensuring the convergence II: Restrictions on paths} 
\label{secA.4}

For a partition $\P$ satisfying the above restriction, the regulated
volume operator is given by
\be\label{volada}
\hat{V}_R^{\P}\Psi_\g\ =\ \sum_v\sqrt{\q_{C_v}}\Psi_\g
\ee
where $v$ runs through the set of vertices of $\g$ and $C_v$ is the
cell containing $v$. The resulting vector $\hat{V}_R^{\P}\Psi_\g$ still
depends on the partition $\P$ through path assignments and its limit
as we shrink the cells may still not exist.%
\footnote{In the original derivation \cite{17} of this operator, the
limit is taken \cite{crjl} by setting $\lim\sqrt{...} :=
\sqrt{\lim...}$, the limit under the square root being obtained by
simply setting the holonomies corresponding to the shrunk paths by the
identity matrix.  Note however that diffeomorphism invariant dual
states $\bra S|$ in (\ref{top}) are not sensitive to the shrinking of
paths, so they never disappear as far as the action of such states is
concerned. This means that if one wishes to use the same topology
(\ref{top}) as in the definition of the Hamiltonian constraint
operator, the final limit would be different from the one reported in
\cite{17}.}
Indeed, for a given vertex $v$ and generic path assignment in $\P$,
the vector $\sqrt{\q_{C_v}}\Psi_\g$ contains a component%
\footnote{Here `component' refers to the decomposition of $\cyl(\Ab)$
into orthogonal subspaces associated to distinct graphs given by 
spin-networks \cite{10,11,9,1}.}
orthogonal to $\sqrt{\q_{C_{v'}}}\Psi_\g$ corresponding to any vertex
$v'\not= v$ of $\g$.  So the problem of convergence reduces to the
convergence of each $\sqrt{\q_{C_v}}\Psi_\g$ as we shrink $C_v$ to $v$
separately for every vertex $v$. But if we shrink $C_v$ to a cell
${C'}_v$ containing $v$ then for a generic paths assignment, the norm
$\|\sqrt{\q_{C_{v}}}\Psi_\g- \sqrt{\q_{{C'}_{v}}}\Psi_\g\|$ is non-zero,
{\it independently} of ${C'}_v$.  The main reason is that
$\sqrt{\q_{C_{v}}}\Psi_\g$ has a non-zero component on a graph $\g_1$
obtained from $\g$ by introducing new vertices made at the
intersection points $w_I$ with $\pC$ and adding new edges, the paths
$p_{w_Ix_0}$. The graph ${\g'}_1$ corresponding to
$\sqrt{\q_{{C'}_{v}}}\Psi_\g$ is (generically) different then $\g_1$.
Moreover, the projection of $\sqrt{\q_{{C'}_{v}}}\Psi_\g$ on the space
corresponding to ${\g'}_1$ is, generically, diffeomorphism equivalent
to the projection of $\sqrt{\q_{{C}_{v}}}\Psi_\g$ on the space
corresponding to ${\g}_1$.  Hence shrinking does not affect the norm
of the projection. All this makes the sequence of vectors
$V_{\P}\Psi_\g$ highly non-convergent as $\P$ is refined.

These problems obviously disappear if, given a graph $\g$, we choose
a paths assignment such that for the above graphs we have
\be \label{grpr} {\g}_1\ =\ {\g'}_1\ =\ \g.  \ee
for every cell. Such an assignment does exist and was pointed out to
us by De Pietri \cite{18}. It is defined as follows.  Denote the edges
of $\g$ that meet at $v$ by $e_I$, $I=1,...,k$, and orient them to be
outgoing. Denote by $e^-_I$ the segment of $e_I$ contained inside
$C_v$ and  by $e^+_I$ the remaining segment of $e_I$.  To each
intersection point $w_I\in e_I\cap\pC_v$ set $p_{w_Iv}\ :=\ e^{-}_I $
and to a point $z\in \pC_v$ which is not intersected by an edge,
assign any path $p_{zv}$ (such that the previous regularity
assumption (\ref{pr}) is satisfied).

With these restrictions on the partition $\P$, the regulator can be
removed unambiguously. We will evaluate the limit in the next two
sub-sections. For the present, we will just comment on the similarity
between the restrictions on partitions used in the two regularization
schemes: Given a cell containing a vertex $v$, making all the paths
begin right at $v$ is the `external' regularization counter part of
the `internal' regularization assumption that the 2-surfaces $S_a$ in
$C$ intersect exactly at $v$.

To conclude this discussion let us note that the issue of path
assignments did not arise in the `internal' regularization because we
worked not on the space $\Ab/\Gb$ of generalized connections modulo
gauge transformations but on the space $\Ab$ of generalized
connections themselves. The loop representation used in \cite{17}, on
the other hand, deals only with $\Ab/\Gb$ and the extra paths become
necessary to ensure gauge invariance of smeared triads. In the
connection representation, there is a choice: We could have avoided
the issue of path dependence entirely by carrying out the `external'
regularization on $\Ab$. We did not do so to keep as close to the
treatment of \cite{17} as possible.
 
\subsection{The evaluation} \label{secA.5}

Let us now return to (\ref{|.|}) and investigate the operator
$\Tr(\E_{\pC, f}(z_1)\E_{\pC, f}(z_2)\E_{\pC, f}(z_3))$. One can show
that only the terms $w_1\not=w_2\not=w_3\not=w_1$ contribute to the
sum which is relevant for the self adjointness of the entries of
$|...|$. To see this, note that, in terms of the segments $e^-_I,
e^+_I$ of $\g$ that meet at an intersection point $w_I\in\pC$, we
have:
\be 
J_{w_I}^{i}\ =\ {1\over 2}(J_{e^+_I}^{i}\ -\ J_{e^-_I}^{i}).  
\ee 
In particular, restricted to $\cyl_\g(\Ab)$, the operators associated
with $w_I$ commute,
\be [J_{w_I^i},J_{w_I^j}]\ =\ 0 \, .  \ee 
Hence the term corresponding to $w_I=w_J$ 
\be
\Tr\big(\tau_i\tau_jU_{w_Iw_K}\tau_kU_{w_Kw_I}\big)
J_{w_K}^{k}J_{w_I}^{j}J_{w_I}^{i}\ \ee 
vanishes because the trace is antisymmetric in $i,j$.

Let us now suppose that the partition $\P$ satisfies the restrictions
of the last two sub-sections. Using De Pietri's path assignment,
denote the parallel transports along $e_I$, $e^-_I$ and $e^+_I$ by
$U_I$, $U^-_I$ and $U^+_I$ respectively. The action of $J_{w_I}^{i}$
on a cylindrical function $\Psi_\g(A)=\psi(U_1(A),...,U_n(A))$ is
given by
\be J_{w_I}^{i}\Psi_\g(A)\ =\ (U_I^+(A)\tau_i
U_I^-(A))^A_B{\partial \psi\over \partial {U_I}^A_B}.  \ee 
So that Eq (\ref{|.|}) now reads 
\ba\label{|..|} 
& &\Tr(\E_{\pC, f}(z_1)\E_{\pC, f}(z_2)\E_{\pC, f}(z_3))\Psi_\g(A)\ 
\nonumber\\
&=&\ 
\sum_{w_I\not=w_J\not=w_K\not=w_I\in\pC}
f(z_1, w_I)f(z_2, w_J)f(z_3, w_K)\nonumber\\
& & \times \Tr\Big(\tau_i U^-_I(U^-_J)^{-1} \tau_j
U^-_j(U^-_K)^{-1}\tau_k U^-_K (U^-_I)^{-1} \Big)
(U^+_I\tau_i U^-_I)^A_B (U^+_J\tau_jU^-_J)^C_D (U^+_K\tau_kU^-_K)^E_F
\nonumber\\ 
& &\times  {\partial^3\over \partial {U_I}^A_B \partial{U_J}^C_D
\partial {U_K}^E_F} \psi(U_1(A),...,U_n(A)).  \ea

The evaluation of the right hand side is considerably simplified by
the following property which is a consequence of (\ref{grpr}).  The
operator $\Tr(\E_{\pC, f}(z_1)\E_{\pC, f}(z_2) \E_{\pC, f}(z_3))$
defines a map
\be \cyl^{(3)}_\g(\Ab)\ \rightarrow\ \cyl^{(0)}_{\g_1}(\Ab) \ee 
where $\g_1$ is the graph obtained from $\g$ by splitting the edges at
the points where they intersect $\pC$.  However, since the operator is
gauge invariant, it preserves the invariance of $\Psi_\g$ with respect
to the gauge transformations at the points of $\pC$.  Therefore,
\be 
\Tr \big(\E_{\pC, f}(z_1)\E_{\pC, f}(z_2) \E_{\pC, f}(z_3)\big)\ :\ 
\cyl^{(3)}_\g(\Ab)\ \rightarrow\ \cyl^{(0)}_\g(\Ab).  \ee 
That means that the right hand side of (\ref{|..|}) is in fact a
function of the entire parallel transports $U_I=U_I^+U_I^-$ so that
one may make the substitution $U^-_I=1$ and $U^+_I=U_I$ to derive its
general form. The result for the action of the operator in the space
$\Cyl_\g(\Ab)$ is 
\be\label{|...|} \Tr \big(\E_{\pC, f}(z_1)\E_{\pC, f}(z_2)\E_{\pC,
f}(z_3)\big)  = \ -{1\over 4}\sum_{I\not=J\not=K\not=I}
f(z_1, w_I)f(z_2,w_J)f(z_3, w_K) \e_{ijk}
J_{ve_I}^{i}J_{ve_J}^{j}J_{ve_K}^{k}, \ee
where $I,J,K$ run through the set labelling the edges of $\g$
intersecting at $v$ and, given $I$, $w_I$ is the intersection point of
$e_I$ with $\pC$.

\subsection{The integral} \label{secA.7} 

We can now turn to the quantum version of the full formula
(\ref{RSdet'}). Thus, we want to evaluate
\be \q_C\, \Psi_\g\ =\ \frac{1}{12}\, \Big(\int_{(\pC)^3}\, \Tr
\big(\E_{\pC, f}(z_1) \E_{\pC, f}(z_2) \E_{\pC, f}(z_3)\,\big)
d^2z_1\, d^2z_2\, d^2z_3\,\,\Big)\,
\Psi_\g \ee
for $\Psi_\g\in\cyl^{(3)}_\g(\Ab)$. The operator $\E_{\pC,
f}(z_1)\E_{\pC, f}(z_2)\E_{\pC, f}(z_3)$ is defined by (\ref{|...|})
in a subspace $\cyl^{(3)}_\g$ for every $\g$.  It is easy to see that
the result agrees on any $\cyl^{(3)}_\g(\Ab)\cap\cyl^{(3)}_{\g'}(\Ab)$
so the operator is consistently defined in $\cyl(\Ab)$. Due to the
self-adjointness of the $J$ operators and the commutativity of
operators associated with different edges, the total operator is
essentially self-adjoint in $\cyl^{(3)}(\Ab)$. Hence, it is meaningful
to take its absolute value.

Apriori, the integral of an operator-valued function may not be well
defined. However, in our case, the only $z_1,z_2,z_3$ dependence of the
integrand comes from the functions $f$ in (\ref{|...|}).  Given a
graph $\g$ and a cell $C$ choose $f$s such that for every two edges
$e_I\not=e_J$ intersecting $C$, the supports $\Delta_I$ and $\Delta_J$
of the functions $f(w_I,.)$ and $f(w_J,.)$ satisfy
$\Delta_I\cup\Delta_J=\emptyset$. Then in (\ref{|...|})
\be
\int_{(\pC)^3}\, |\sum_{I,J,K}...|\,d^6z\,\, \Psi_\g\ =\
\sum_{I,J,K}\int_{\Delta_I}\int_{\Delta_J}\int_{\Delta_K}
f(z_1, w_I) f(z_2,w_J)f(z_3,w_K)\,d^6z\,\, |...|\,\,\Psi_\g 
\ee
where the last factor $|...|$ is $z$ independent and where, due to the
normalization property, the $f$'s integrate out to $1$.  Finally, for a
cylindrical function $\Psi_\g$ compatible with a graph $\g$ and a cell
$C_v$ containing a vertex $v$,
\be\label{6.24} 
\q_{C_v}\Psi_\g \ =\ {1\over 48}\sum_{I\not=J\not=K\not=I}
|\e_{ijk}J_{ve_I}^iJ_{ve_J}^jJ_{ve_K}^k|\Psi_\g \ee 
where $I,J,K$ label the edges at $v$. Now the situation is the same as
that we encountered in Section \ref{sec3}: the formula is manifestly
invariant with respect to the shrinking of of the cells $C_v$.  Hence,
for any partition $\P$ which satisfies the restrictions stated in the
previous sub-sections, the volume operator is given by
\be\label{volim} 
V_R\,\Psi_\g\ \ =\ \sum_{v} \sqrt{\q_v}\, \Psi_\g \, ,\ee 
where the sum extends over all the vertices of the graph. This is the
formula that was reported in \cite{15}. The final expression is
insensitive to the `intrinsic' structure of the graph at its vertices 
and depends only on the `extrinsic' structure which can be registered 
at the boundary of sufficiently small cells surrounding the vertices.

\end{document}